\documentclass[12pt]{article}
\usepackage{amsmath}
\usepackage{epsfig}
\usepackage{amssymb,amsfonts}

\advance\voffset by -2.0cm
\advance\hoffset by -1.25cm
\def\be{\begin{equation}}
\def\ee{\end{equation}}
\def\ba{\begin{eqnarray}}
\def\ea{\end{eqnarray}}

\newcommand{\E}{{\cal E}}
\newcommand{\G}{{\cal G}}

\newcommand{\ZZ}{\mathbb{Z}}
\newcommand{\BC}{\mathbb{C}}
\newcommand{\IP}{\mathbb{P}}

\newcommand{\ie}{{\it i.e.~}}

\newcommand{\Zop}{\mathbb{Z}}

\newcommand{\half}{\frac{1}{2}}

\begin{document}

\vspace*{-1.5cm}
\thispagestyle{empty}
\begin{flushright}
hep-th/0603196
\end{flushright}
\vspace*{2.5cm}

\begin{center}
{\Large 
{\bf Matrix factorisations and D-branes on K3}}
\vspace{2.0cm}

{\large Ilka Brunner}%
\footnote{{\tt E-mail: brunner@itp.phys.ethz.ch}}, 
{\large Matthias R.\ Gaberdiel}%
\footnote{{\tt E-mail: gaberdiel@itp.phys.ethz.ch}} 
{\large and}
{\large Christoph A.\ Keller}%
\footnote{{\tt E-mail: kellerc@itp.phys.ethz.ch}} 

\vspace*{0.5cm}

Institut f\"ur Theoretische Physik \\
ETH Z\"urich\\
8093 Z\"urich, Switzerland\\
\vspace*{2cm}

{\bf Abstract} 
\end{center}

\noindent D-branes on K3 are analysed from three different points
of view. For deformations of hypersurfaces in weighted projected 
space we use geometrical methods as well as matrix factorisation
techniques. Furthermore, we study the D-branes on the $T^4/\Zop_4$
orbifold line in conformal field theory. The behaviour of the D-branes
under deformations of the bulk theory are studied in detail, and 
good agreement between the different descriptions is found.

\newpage
\renewcommand{\theequation}{\arabic{section}.\arabic{equation}}


\section{Introduction}

D-branes in models with $N=(2,2)$ worldsheet supersymmetry have been
discussed in recent years from various points of view. Prominent
examples for such theories include non-linear sigma-models whose
targets are K\"ahler manifolds, Gepner models and Landau-Ginzburg
models. Boundary conditions can be formulated in terms of the
$N=(2,2)$ supersymmetry algebra, and one is usually 
interested in boundary conditions that preserve half of the
supersymmetry. As is well known, there are two different classes of
such supersymmetry preserving boundary conditions, which are related
by mirror symmetry and are called A-type and B-type. In the non-linear
sigma-model, A-type boundary conditions correspond to D-branes
wrapping (special) Lagrangian cycles, whereas B-type boundary
conditions describe holomorphic branes \cite{OOY}. In 
the Gepner model, which provides a rational conformal field theory
description in the small volume regime of certain Calabi-Yau
compactifications, A-type and B-type D-branes can be constructed as
explicit boundary states with appropriate gluing conditions for the
generators of the symmetry algebra. Finally, in the Landau-Ginzburg
model, A-type D-branes correspond to Lagrangian submanifolds that are
mapped by the superpotential to straight lines \cite{HIV}, whereas
B-type D-branes can be described in terms of matrix factorisations of
the superpotential. The study of matrix factorisations was initiated
in this context by Kontsevich, who proposed that there is a B-type
D-brane for any factorisation of the superpotential 
$W(x_i)=E(x_i)J(x_i)$ in terms of matrices $E$ and $J$ with
homogeneous polynomial entries. One then associates a BRST operator of
the form  
\be
Q=\left(\begin{array}{cc}0&J\\E&0\end{array}\right) 
\ee 
to this factorisation and determines the (topological) open string
spectrum as the BRST cohomology of this operator 
\cite{Kapustin:2002bi,BHLS,Kapustin:2003,Kapustin:2003rc,L,HL}.

In this paper, we shall analyse supersymmetric D-branes on K3 surfaces
using matrix factorisation and conformal field theory techniques. 
K3 surfaces are special since they actually preserve $N=(4,4)$
supersymmetry, not just $N=(2,2)$. The relation between the
geometrical and conformal field theory description of closed strings
on K3 has been analysed in some detail in \cite{Nahm:1999ps}; here
we shall concentrate mainly on open strings. A-type and B-type
boundary conditions can be formulated as usual once a particular $N=2$
subalgebra is chosen. Geometrically, all supersymmetric D-branes are
holomorphic with respect to some complex structure. For a given
complex structure, on the other hand, not all of these D-branes are
holomorphic and the spectrum of holomorphic D-branes thus depends on
the actual point in moduli space. One expects to find a D0 and D4
branes at any  point in moduli space, but the rank of the Picard
lattice, that determines the number of holomorphic curves and hence D2
branes, varies.      

An explicit conformal field theory description of string theory on K3
is only available at rather specific points in the moduli space, but
the matrix factorisation description is, in principle, available for a
much larger subspace of the moduli space. Thus the latter approach is
very well suited to study the spectrum of B-type D-branes for generic
points in the moduli space. In this paper we shall study some aspects
of matrix factorisations for K3s that can be described as hypersurfaces 
in weighted projective space, where we restrict to the case of
Fermat type polynomials and their perturbations. 
At the maximally symmetric point, string theory on this surface has a
conformal field theory description in terms of a Gepner model.  

At this Gepner point we can easily construct factorisations by
tensoring the usual single factor and the permutation factorisations
together.\footnote{Unlike the situation for 
$3d$ Calabi-Yau's \cite{CFG} these factorisations do not
in general account for all B-type RR charges. This is a consequence of
the fact that for K3 the middle dimensional cycles are
2-dimensional. Given the relation between matrix factorisations and
geometry, one should thus not expect to obtain the charges of all
holomorphic 2-cycles by these constructions.} We can 
then analyse how these B-type D-branes can be deformed as one perturbs
the superpotential. For certain factorisations (in particular, the
tensor product factorisation and single transposition factorisations)
we will be able to show that they can be extended over the whole
moduli space of deformations. This can be done very explicitly by
writing down the  corresponding factorisations for arbitrarily
deformed superpotentials (see section~3). This result has a very nice
geometrical interpretation: these factorisations account precisely for
those holomorphic D2-branes that come from the embedding  space or
from resolving singular points of the embedded manifold, and are 
generically present for K3s that arise as hypersurfaces in weighted
projective space.    

On the other hand, we can also prove that certain factorisations 
{\it cannot} be deformed. This can be shown by studying the
infinitesimal deformations following \cite{Hori:2004,Ashok:2004xq}. 
In particular,
we shall show that a necessary condition for a deformation not to 
be obstructed is that the brane is uncharged under the RR ground state
corresponding to the deforming polynomial. The fact that
factorisations are generically obstructed is also in good 
agreement with the geometrical results.
\smallskip

In order to compare these results with what can be analysed in
conformal field theory we can make use of the fact that there is an 
interesting subspace of the moduli space of such K3 surfaces for which
we have an explicit conformal field theory description. Indeed, the
deformations of the quartic surface in $\IP_3$ by two particular bulk
fields 
\be\label{orbline}
x_1^4 + x_2^4 + x_3^4 + x_4^4 + a x_1^2 x_2^2 + b x_3^2 x_4^2=0 
\ee
describes \cite{Wend05} the 2-parameter space of toroidal $\ZZ_4$
orbifold K3s. For this subspace of the moduli space one should thus be
able to relate the different matrix factorisations with explicit
conformal field theory constructions of D-branes; this will be done in
section~4. 

In the orbifold description it is straightforward to see that the
space of B-type RR charges is $22$-dimensional for any choice of the
orbifold parameters $a$ and $b$,\footnote{The relation between the two
theories involves mirror symmetry. These D-branes are therefore A-type
from the point of view of the orbifold theory.} and it is in principle
not difficult to construct the relevant D-branes in the orbifold
conformal field theory. On the matrix factorisation side it is 
likewise not difficult to construct the relevant $22$ factorisations 
that account for all of these charges at 
the Gepner point (where $a=b=0$). However, we can show that not all of
them can be extended to arbitrary $a,b$. In fact, given our general
results about obstructions, it is clear that for the  factorisations
that are charged under the RR ground states corresponding to $x_1^2
x_2^2$ or $x_3^2 x_4^2$ this will not be possible. Furthermore, we can
show that this is the only real obstruction: we have identified a set 
of factorisations that account for $20$ RR charges and that can be 
extended for arbitrary $a$ and $b$.

This apparent obstruction has in fact a very nice interpretation in
terms of the orbifold conformal field theory. At least some of the
relevant D-branes that carry these charges stretch diagonally across
the two $T^2$s at $45$ degrees. Varying the parameters $a$ and $b$
corresponds then to changing the  radii of the two tori (as well as
switching on a $B$-field). The structure of the $45$ degree
D-brane\footnote{One may also consider modifying the angle with which
the D-brane stretches across the tori. However, the resulting D-brane
will then typically not satisfy the correct $N=2$ gluing conditions
any more.} then depends crucially on the relative radii: if their
ratio is rational, the brane will have finite length, but it will wind
infinitely many times around the torus if the ratio is irrational 
\cite{DougHull}. Thus the boundary state depends in a very
discontinuous manner on the parameters of the closed string
theory. (This is also familiar from the analysis of the $N=0$ and
$N=1$ D-branes on a single circle
\cite{Gaberdiel:2001zq,Janik,Friedan}.) This explains why the 
corresponding matrix factorisation description cannot depend on the
deformation parameters in a simple analytic way.
\smallskip 

The paper is organised as follows. In section \ref{geom} we review 
some background material about D-branes on K3 surfaces from a
geometric point of view. In particular, we describe how the rank of
the Picard lattice is always at least $1$ for K3 surfaces that are
embedded in  weighted projective space. In section \ref{matrix} we
briefly review the matrix factorisation approach to D-branes in Landau
Ginzburg models. We then discuss the behaviour of D-branes under bulk 
perturbations in this language, and explain, in particular, how the 
generic rank of the Picard lattice can be understood from this
perspective. We also study the quartic in $\IP_3$ and the above
deformations in detail. Finally, in section \ref{CFT} we discuss the
orbifold theory $T^4/\ZZ_4$ and its D-branes. We explain some parts of
the correspondence between the boundary states of the orbifold theory
and the matrix factorisations of the Landau-Ginzburg
description. Finally we study their deformations in the orbifold
theory and explain why certain deformations are obstructed. We have
included an appendix in which some of the more technical aspects of
the orbifold description and its dictionary to the Gepner model are
explained in detail.


\section{BPS D-branes on K3 surfaces}\label{geom}

Let us begin by explaining some generalities about D-branes on K3 
surfaces \cite{OOY,BVS,YZ}.  On K3 we are in the special situation
that there is extended $N=(4,4)$ supersymmetry. The $N=4$ algebra is
an extension of the usual $N=2$ superconformal algebra, where the 
$u(1)$ current of the $N=2$ theory is enhanced to an
$\widehat{su}(2)_1$ algebra; the additional generators are the
spectral flow operators (by one unit), which have conformal weight $1$
for $c=6$.  

{}From the point of view of the extended $N=(4,4)$ symmetry there is
therefore some freedom in how to choose the $u(1)$ generator of the
$N=2$ algebra inside the $\widehat{su}(2)_1$ algebra of the $N=4$
algebra. This is precisely the freedom of choosing a Cartan torus for
the $SU(2)$ group. Each $N=2$ subalgebra determines uniquely an $u(1)$
subalgebra of the $\widehat{su}(2)_1$, but the converse is not true
\cite{Wend05,Huy}.
Once we have identified in addition a particular $N=(2,2)$ subalgebra,
we can formulate A and B type boundary conditions as usual. However,
it is clear that the  distinction between A-type and B-type branes
depends on the choice of the particular $N=(2,2)$.  

Mirror symmetry corresponds algebraically to flipping the sign of the
$u(1)$ current of the left moving supersymmetry algebra. Obviously,
this operation requires that a particular $N=(2,2)$ structure has been
picked.  The mirror operation can then be viewed as a rotation of the
Cartan torus (for the left movers).

Geometrically, a $K3$ surface $S$ is a hyperkaehler manifold with
$H^2(S,\ZZ)=22$. With respect to the usual intersection product, the
resulting lattice is even and self-dual, and has signature
$(+)^3,(-)^{19}$. A hyperkaehler structure is determined by the
positive 3-plane spanned by the periods of the three hyperkaehler
forms in that lattice. Once a compatible complex structure is chosen,
this three-plane has an orthogonal decomposition into the line
generated by $\omega$ (the Kaehler form), and the plane spanned by the
real and imaginary components of the holomorphic $2$-form
$\Omega=x+iy$. A change of complex structure amounts to rotating the
2-plane spanned by the vectors $x,y$. In the context of string theory,
the moduli space contains in addition the B-field, and the full 
moduli space takes the form of a Grassmannian 
parametrising $4$-planes in $\mathbb{R}^{4,20}$. A decomposition of
the positive 4-plane into two orthogonal 2-planes then amounts to
fixing the complex structure, a Kaehler class and a B-field. In
\cite{Dij}, the four 2-forms (three Hyperkaehler forms and the
B-field) have been combined into a single quaternionic
$2$-form. Mirror symmetry, which interchanges the complex structure
with the complexified Kaehler structure, acts in this language as a
quaternionic rotation of the positive 4-plane.

Comparing with the conformal field theory description, the choice of a
decomposition of the positive 4-plane into two perpendicular
2-planes amounts to the choice of an $N=(2,2)$ subalgebra inside the 
$N=(4,4)$. The two $SU(2)$ enhancing the $N=(2,2)$ to $N=(4,4)$ can be
understood as the freedom to rotate the two 2-planes.  

Geometrically, B-type D-branes correspond to holomorphic branes,
whereas A-type branes wrap (special) Lagrangian submanifolds. In the
case of $K3$, B-type branes can have dimension $0,2,4$, whereas A-type
branes are always 2-dimensional. Some of the $22$ 2-cycles will thus
be wrapped by A-type branes, and some by B-type branes, but the
decomposition into A-type and B-type branes depends, of course, on the
chosen complex structure. For example, the quaternionic rotation that
induces mirror symmetry exchanges holomorphic and Lagrangian cycles.
The action of mirror symmetry on the D-branes can also be understood
from the point of view of \cite{SYZ}, where mirror symmetry was
formulated for elliptic fibrations with a section as T-duality on the
fiber.  In the K3 context, this point of view has been used to extend
mirror symmetry to the open string sector in \cite{BBS}. Homological
mirror symmetry has been proven for the quartic surface in \cite{Sei}.

\subsection{B-type branes and the Picard lattice}

As we have explained above, supersymmetric 2-cycles on K3 are
holomorphic curves with respect to {\it some} complex
structure. If a 2-cycle is holomorphic with respect to a given complex
structure, it can be wrapped by a D-brane that is B-type with respect
to the corresponding $N=(2,2)$ subalgebra. In the following we review 
some background material from \cite{Asp}. 


The $2$-cycles are naturally elements of $H_2(S,\ZZ)$, or, using
duality, of $H^2(S,\ZZ)$. Holomorphicity imposes that the dual 
$2$-form is in fact in $H^{1,1}(S)$, and the Picard lattice is thus 
\be
{\rm Pic}(S) = H^2(S,\ZZ) \cap H^{1,1}(S) \ .
\ee
The rank of the Picard lattice is usually denoted by
$\rho$. Generically, K3 surfaces will have $\rho=0$, meaning that no
B-type 2-branes are compatible with the given holomorphic
structure. However, in this paper we will always consider special
geometric points at which the rank of the Picard lattice is enhanced
or even maximal.  

We are particularly interested in the case where $S$ is a hypersurface 
described by a Fermat polynomial in a weighted projective space. In
such a case, there is a correspondence \cite{phases} between the  
non-linear sigma model on the hypersurface and the Landau Ginzburg
model with a superpotential that formally equals the polynomial
appearing in the hypersurface equation. To be more precise, the
hypersurface in $\IP_{w_1,w_2,w_3,w_4}[H]$, with $H=\sum w_i$ 
\be
x_1^{k_1+2} + x_2^{k_2+2} + x_3^{k_3+2} + x_4^{k_4+2} = 0\ ,
\ee
where $k_i+2=H/w_i$, $H={\rm lcm} \{ k_i+2 \}$, corresponds to the
Landau Ginzburg orbifold model with superpotential 
\be \label{weightedW}
W=x_1^{k_1+2} + x_2^{k_2+2} + x_3^{k_3+2} + x_4^{k_4+2}\ .
\ee
In this equation the $x_i$ denote chiral superfields of charge 
$q_L=q_R=1/(k_i+2)$; our notation will not distinguish between the
chiral fields of the Landau-Ginzburg model and the coordinates of the
projective space in the geometric description. The orbifold $\ZZ_H$
acts by phase multiplication on the chiral superfields 
$x_i\mapsto e^{\frac{2\pi i}{k_i+2}} x_i$; this
orbifold projects onto integer $U(1)$ charges of the theory.
Altogether, there are 14 different examples
corresponding to Fermat polynomials in weighted projective space, which
we list in table~1. These models also have a description in terms 
of rational conformal field theory, namely as the tensor product of $4$
$N=2$ minimal models at levels $k_i$, modulo an integer charge
projection $\ZZ_H$. In terms of conformal field theory, it is
straightforward to see that the space of B-type RR charges is 
$22$-dimensional for each of these models. By standard conformal field
theory arguments one should therefore expect that the corresponding
B-type D-branes exist, and thus that the rank of the Picard lattice is
maximal for all of these points.

\begin{table}[ht]\begin{center}
\begin{tabular}{|l|c|c|c|c|}
\hline
 Projective space&$W(x_1,x_2,x_3,x_4)$& Minimal model   &  RS & RS,ST \\
\hline
$\mathbb{P}_{(1,1,1,1)}[4]$& $x_1^4+x_2^4+x_3^4+x_4^4$ & $(2,2,2,2)$ &
3 &3 \\ 
$\mathbb{P}_{(1,1,1,3)}[6]$ &$x_1^6+x_2^6+x_3^6+x_4^2$& $(4,4,4,0)$
&3&3  \\ 
$\mathbb{P}_{(1,1,2,2)}[6]$ &$x_1^6+x_2^6+x_3^3+x_4^3$& $(4,4,1,1)$
&4&6\\ 
$\mathbb{P}_{(1,1,2,4)}[8]$ &$x_1^8+x_2^8+x_3^4+x_4^2$& $(6,6,2,0)$
&4&5\\ 
$\mathbb{P}_{(1,2,2,5)}[10]$ &$x_1^{10}+x_2^5+x_3^5+x_4^2$&
$(8,3,3,0)$ &4&8 \\ 
$\mathbb{P}_{(1,1,4,6)}[12]$&$x_1^{12}+x_2^{12}+x_3^3+x_4^2$&
$(10,10,1,0)$ &4&4 \\ 
$\mathbb{P}_{(1,2,3,6)}[12]$&$x_1^{12}+x_2^6+x_3^4+x_4^2$&
$(10,4,2,0)$ &6&9 \\ 
$\mathbb{P}_{(1,3,4,4)}[12]$&$x_1^{12}+x_2^4+x_3^3+x_4^3$&
$(10,2,1,1)$ &6&12 \\ 
$\mathbb{P}_{(2,3,3,4)}[12]$ &$x_1^6+x_2^4+x_3^4+x_4^3$& $(4,2,2,1)$
&6&14 \\ 
$\mathbb{P}_{(1,2,6,9)}[18]$ &$x_1^{18}+x_2^9+x_3^3+x_4^2$&
$(16,7,1,0)$ &6&8 \\ 
$\mathbb{P}_{(1,4,5,10)}[20]$ &$x_1^{20}+x_2^5+x_3^4+x_4^2$&
$(18,3,2,0)$ &8&12 \\ 
$\mathbb{P}_{(1,3,8,12)}[24]$ &$x_1^{24}+x_2^8+x_3^3+x_4^2$&
$(22,6,1,0)$ &8&10 \\ 
$\mathbb{P}_{(2,3,10,15)}[30]$ &$x_1^{15}+x_2^{10}+x_3^3+x_4^2$&
$(13,8,1,0)$ &10&14  \\ 
$\mathbb{P}_{(1,6,14,21)}[42]$ &$x_1^{42}+x_2^7+x_3^3+x_4^2$&
$(40,5,1,0)$ &12&12 \\ 

\hline
\end{tabular}
\end{center}
\caption{The $14$ different K3 that correspond to Fermat polynomials
in weighted projective space. The last two entries denote the rank
of the charge lattice spanned by RS-branes and by RS and single
transposition branes. As argued below in section 3.3.2, the last entry
minus two (two charges correspond to D0- and D4-branes) should equal
the rank of the Picard lattice at a generic point in the complex
structure moduli space of the corresponding surface.} 
\label{t:CL}
\end{table}

The condition that $S$ can be written as a hypersurface in weighted
projective space imposes constraints on the allowed complex structure
deformations, and thus increases the generic rank of the Picard
lattice \cite{Asp}. For example, in the case of the quartic in  
$\IP_3$  
\be
x_1^4+x_2^4+x_3^4 + x_4^4+ \dots=0
\ee
there is at least one holomorphic curve at any point in the complex 
structure moduli space, namely the intersection of the quartic
polynomial with any hyperplane. This phenomenon generalises
immediately to all hypersurface equations in weighted projective
space, where one can always consider the intersection with a
hyperplane. In some examples the generic rank of the Picard lattice
may be enhanced even further. Consider for example the model 
$\IP_{(1,1,2,2)}[6]$. The embedding weighted projective space 
has a $\ZZ_2$ orbifold singularity with fixed point $(0,0,x_3,x_4)$.  
The $\ZZ_2$ singularity is resolved by an exceptional $\IP_1$. It
intersects with the hypersurface in the $3$ points that are defined by
the equations $x_1=x_2=0$ and $x_3^3+x_4^3=0$. This enhances the rank
of the generic Picard lattice by $3$. Altogether, the rank of the Picard
lattice is therefore $4$ at generic points in the complex structure moduli
space for this example. Note that there is one 2-cycle that
is inherited from the embedding space: it corresponds to the 
combination of the three spheres, which is invariant under 
$x_3\mapsto \exp(2\pi i/3) x_3$, which permutes the 3 singular 
points on the hypersurface. There are therefore $4$ different brane
charges, two D2, the D0 and D4 that the hypersurface inherits directly
from the embedding space.

More generally, whenever two weights have a greatest common divisor
$m$ by which the other two weights are not divisible, the embedding
projective space acquires an orbifold singularity  which locally has
the form $\BC^2/\ZZ_m$.  Its resolution requires $m-1$ $\IP_1$s  
whose intersection pattern is given by the $A_{m-1}$ Dynkin diagram. 
This means that any such singularity  contributes $m-1$ 2-brane
charges to the charge lattice that can be obtained by pull back from
the embedding projective space. To determine the contribution to the
Picard lattice of the hypersurface, one has to take into account that,
as in the example above, the hypersurface might intersect the
exceptional set in several points. Each of them gives a contribution
to the Picard lattice. We will interpret these general
features of the Picard lattice from
the matrix factorisation point of view in section \ref{matrix}.

\subsection{The orbifold line}

Generically, the points in moduli space where a conformal field theory
description is known are isolated. For example, for the above theories
we only have a conformal field theory description (namely a Gepner
model) for the unperturbed superpotential. There is, however, one
interesting exception to this: since the $(2)^4$ model is in fact
equivalent to the $\Zop_4$ toroidal orbifold \cite{Nahm:1999ps}, there
is a two-parameter family of orbifold theories all of which describe 
K3. The corresponding subspace of the moduli space has recently been 
identified to be \cite{Wend05} 
\be\label{veryattractive}
x_1^4+x_2^4+x_3^4+x_4^4 + a x_1^2 x_2^2 + bx_3^3 x_4^2 =0 \ .
\ee
The orbifold theory will be described in more detail in section~4; the
detailed mapping between $a$ and $b$ and the relevant parameters of
the orbifold theory was given in \cite{GS,Wend05}.

For this subspace of the moduli space we therefore have a good
understanding of both the conformal field theory and the matrix
factorisation approach. We should thus be able to compare the 
results from both points of view. The matrix factorisation description
will be given in the following section, where we will in particular
show that certain D-branes are obstructed against modifying the bulk
parameters $a$ and $b$. In section~4 we will identify the
corresponding boundary states in the orbifold conformal field theory
and reproduce these obstructions also from that point of view.

\section{The matrix factorisation point of view}\label{matrix}
\setcounter{equation}{0}

In this section we shall analyse the above theories from the matrix
factorisation perspective. This approach was proposed in unpublished
form by Kontsevich, and the physical interpretation of it was given in 
\cite{Kapustin:2002bi,BHLS,Kapustin:2003,Kapustin:2003rc,L,HL}, for a
review see also \cite{HWs}. We shall first collect very briefly some
basic facts about matrix factorisations that we shall need later on.

\subsection{Fundamentals}

Kontsevich has proposed that D-branes in a Landau-Ginzburg 
models are given by matrix factorisations of the superpotential,
\be
Q^2=W\cdot \mathbf{1}\ ,
\ee
where $Q$ is a square matrix with polynomial entries that satisfies
\be 
\sigma Q + Q\sigma=0\ .
\ee
If we choose the grading operator $\sigma$ to be diagonal, $Q$ is of
the form 
\be
Q=\left(\begin{array}{cc}0&J\\E&0\end{array}\right)\quad
\textrm{with}\quad EJ=JE=W\cdot\mathbf{1}\ .
\ee
Two factorisations $(E,J)$ and $(E',J')$ are considered equivalent 
if they are related by
a similarity transformation with invertible matrices with
polynomial entries,
\be
E'=U_1 E U_2^{-1}\ , \qquad J'=U_2 J U_1^{-1} \ .
\ee
The spectrum of open strings between D-branes determined by
factorisations $(E,J)$ and $(E',J')$ is then given by the cohomology
of the boundary BRST operator $Q$. From a physics point of view, 
the factorisation condition can be derived by varying the
Landau-Ginzburg action and cancelling the boundary terms 
\cite{Kapustin:2002bi,BHLS}. This analysis also confirms that $Q$
is the correct boundary BRST operator. The proposal got further
support by relating the results from the matrix factorisation
perspective with those obtained in conformal field theory; in
particular, this was done for the $N=2$ minimal models in
\cite{BHLS,Kapustin:2003rc,Brunner:2005pq} and for tensor products of 
minimal models in \cite{Brunner:2005fv,Enger:2005jk}. Finally, the
matrix factorsation results for toroidal theories were shown to be in
agreement with geometrical expectations 
\cite{Brunner:2004mt,Govindarajan:2005im,Herbst:2006nn}.

We are particularly interested in Landau-Ginzburg orbifolds of the form
(\ref{weightedW}). In this situation, the orbifold group
$\mathbb{Z}_H$ gives an additional finer grading. This grading
correspond to the choice of a representation $\gamma_M$ such that $Q$
satisfies  
\be
\gamma_M\, Q(\omega^{w_i}x_i)\, \gamma_M^{-1}=Q(x_i) \ , 
\ee
where $\omega=e^{\frac{2\pi i}{H}}$. There are $H$ different choices
for $\gamma_M$ that are labelled by $M$.
 
Given a matrix factorisation $Q$, the charge of the corresponding
D-brane under the RR ground states can be calculated using the
formulas derived in \cite{Kapustin:2003,Walcher:2005,HeLa}. RR ground
states in Landau-Ginzburg orbifolds arise both from the twisted and
untwisted sector and can be counted using the techniques of
\cite{vafaLG}. The RR ground states from the untwisted sector
correspond to polynomials in the Landau-Ginzburg fields and have 
the property that the $U(1)$ charges of the left and right moving part
are equal, $q_L=q_R$. In the simplest case, where the weights do not
have divisors (such as for the quartic in  $\IP_3$), there is exactly
one RR ground states from each twisted sector. 

For the general case, let $n = 0,\ldots, H-1$ label the twisted
sectors. In each sector, consider only the untwisted fields $x_i$ such
that $nw_i/H\in\mathbb{Z}$, and set all twisted fields to zero. Let
$\phi^\alpha_n=\prod_i(x_i)^{\alpha_i}$ be a basis of the
untwisted chiral ring $\mathcal{J}_n=\mathbb{C}[x_i]/\partial W_n$
such that $\sum_i\alpha_iw_i/H=\sum_i(\frac{1}{2}-w_i/H)$. The RR ground
states $\vert n;\alpha\rangle$ are obtained by acting with
$\phi_n^\alpha$ on the unique state $\vert n;0\rangle$.
(For $n=0$ this representation corresponds to (\ref{RRmono}) --- see below.) 
Note that not all of the RR ground states obtained in this way survive
the orbifold projection, which has to be imposed for all twists.
The RR charge of $Q$ with respect to a surviving RR ground state
$\vert n;\alpha\rangle$ is 
given by \cite{Kapustin:2003,Walcher:2005} 
\begin{equation} 
{\rm ch}(Q)(\vert n;\alpha\rangle) 
=\frac{1}{(2\pi i)^{r_n}}\oint dx_1\ldots dx_{r_n}
\frac{\phi_n^\alpha\mathrm{Str}
[\gamma^n_M\partial_1 Q_n\ldots\partial_{r_n}Q_n]}
{\partial_1W_n\ldots\partial_{r_n}W_n}\ .\label{KL}
\end{equation}
Here $r_n$ is the number of untwisted fields,
and $W_n$ and $Q_n$ are the superpotential and the factorisation 
with all twisted fields set to zero. The
supertrace is the trace taken with the grading operator $\sigma$
included, \ie $\mathrm{Str}[\cdot]=\mathrm{tr}[\sigma\cdot]$.  

In the context of the correspondence between Landau-Ginzburg models
and Calabi-Yau manifolds, one would expect that one can associate to
any matrix factorisation an element of the derived category of
coherent sheaves of the Calabi-Yau manifold. The derived category
of coherent sheaves and the category of graded matrix factorisations
have to be equivalent since both are believed to describe the
topological category of B-type branes
\cite{Ashok:2004zb,Walcher:2005}, which is supposed to decouple from
the Kaehler moduli. One way to investigate this correspondence would
be to analyse matrix factorisations in the context of the linear
sigma-model. Since this has not yet been done to date, we will use a
result of Orlov \cite{O}, who established mathematically a
correspondence between the `category of singularities' $D_{Sg}$, and 
the category of matrix factorisations (with the equivalence relations
mentioned above). The category of singularities $D_{Sg}$ is a 
certain quotient of the derived category of coherent sheaves that
depends only on the singularity and would be empty on a smooth
manifold. 

Orlov's equivalence was formulated for  the case of the
un-orbifolded  Landau-Ginzburg model. Given a matrix factorisation
$W=EJ$ one interprets the two factors as maps between projective
modules over the polynomial ring $\BC[x_i]$ --- in our case these
modules are simply $\BC[x_i]^{\oplus n}$ for a factorisation in terms
of $n\times n$ matrices $E,J$. One then associates to a factorisation
the object coker $J$, which naturally lives on $W=0$. This assignment
associates to a single transposition brane in an unorbifolded two 
variable model the set $x_1-\eta x_2=0$. It has been shown
\cite{Brunner:2005fv} that the geometric intersection numbers can be
matched with the intersection numbers derived from matrix
factorisations (as well as with those obtained from permutation
boundary states in conformal field theory). For the case of 
graded matrix factorisations in Landau-Ginzburg orbifolds, the idea is
then that linear factorisations still describe the geometric object
coker $J$ for one choice of the representations $\gamma_M$; the 
D-branes corresponding to the other representations are images of that
brane under the Landau-Ginzburg monodromy. For a number of examples
this assignment has been verified for linear transposition and tensor 
product factorisations in \cite{Ashok:2004zb,Brunner:2005fv}. This was
done by using alternative methods \cite{BDLR} to calculate the large
volume charges corresponding to the branes at the Landau-Ginzburg
point. In this paper, we will use these ideas to guess linear matrix 
factorisations corresponding to certain geometric D-branes.

\subsection{Basic factorisations}

The factorisations we shall mainly consider in this paper can be
obtained as graded tensor products $Q_1\odot Q_2$ 
\cite{Ashok:2004zb,Hori:2004} of two simple classes of
factorisations. The first construction involves a single factor theory
of the form  $W=x^h$, for which we can construct a factorisation as  
\begin{equation}
Q(x)=\left(\begin{array}{cc}
0&x\\
x^{h-1}&0\end{array}\right)\ .
\end{equation}
The branes that correspond \cite{Ashok:2004zb} to the tensor product
of four such  factorisations are the RS D-branes with $L=0$
\cite{Recknagel:1998sb}. It follows from (\ref{KL}) that these branes
do not couple to RR charges in the untwisted sector.  

The other construction involves two factors of the form 
$W(x_1,x_2) = x_1^h+x_2^h$. Let $\eta$ denote an $h^{\rm th}$ root of 
$-1$, then we have the factorisation \cite{Ashok:2004zb}
\begin{equation}
Q_\eta=\left(\begin{array}{cc}
0&(x_1-\eta x_2)\\
\prod_{\eta'\neq\eta}(x_1-\eta'x_2)&0
\end{array}\right)\ .
\end{equation}
It was shown in \cite{Brunner:2005fv} (see also \cite{Enger:2005jk}) 
that the corresponding branes are permutation branes with $L=0$
\cite{Recknagel:2002qq}. More generally, these factorisations can also
be constructed for the case that $h_1$ and $h_2$ have a non-trivial
common factor (but are not equal) \cite{CFG}. The corresponding branes
should then be generalised permutation branes similar to those
considered in \cite{Fredenhagen:2005an}. 

If we tensor this permutation factorisation to two tensor
factorisations, we get a transposition brane, denoted for example by
(34). Once again, it only couples to charges in the twisted
sectors. One can, of course, also consider tensoring with another
permutation brane whenever the divisibility properties of the weights
allow this. We will call the resulting branes double transposition 
branes and denote them by (12)(34), {\it etc}.

\subsection{Deformations} \label{deformations} 

For the following it will also be important to understand how the
B-type D-branes behave under deformations of the complex structure. In 
particular, we will consider deformations of the Landau-Ginzburg
superpotential by suitable quasihomogeneous polynomials of
appropriate weight $V$, $W\mapsto \widehat{W}(\psi)=W+\psi V$, where
$\psi$ denotes the parameter of the deformation. If $Q$ is a
factorisation of $W$, then we ask whether there is
$\widehat{Q}(\psi)=Q+f(\psi) \delta Q$ with $f(0)=0$ such 
that $\widehat{Q}$ is a factorisation of $\widehat{W}$. If such a
$\widehat{Q}(\psi)$ exists (at least in the neighbourhood of $\psi=0$) we
shall say that the D-brane can be extended for the deformation
described by $V$. 

\subsubsection{Global deformations}

There exist some classes of branes that can be extended for all
possible deformations. In particular, this is the case for the tensor
factorisations that correspond to RS branes. In order to see this we
note that we can write any superpotential $\widehat{W}$ as 
\be
\widehat{W}=x_1\, F_1+x_2\, F_2+x_3\, F_3+x_4\, F_4\ ,
\ee
where the $F$ are suitable polynomials. In fact, we have 
\be
F_i = \frac{w_i}{H} \, \frac{\partial \widehat{W}}{\partial x_i}\ .
\ee
We can thus define factorisations that are 4-fold tensor products of
the factorisations $x_i F_i$. These factorisations are the deformations
of the standard tensor factorisations. Indeed, as we approach 
the Gepner point, we have $F_i\rightarrow x_i^{h_i-1}$, and these
factorisations reduce to the tensor branes.
\smallskip

In a similar way, we can extend single transposition branes. 
For definiteness we assume that $w_3=w_4$ and define
\be
L_1=x_1\ , \quad L_2= x_2\ , \qquad L_3=x_3-\alpha x_4\ .
\ee
Inserting $L_1=L_2=L_3=0$ into the superpotential and imposing
$\widehat{W}=0$, one derives an equation of degree $k_3+2$ for
$\alpha$,
\be\label{alpha}
\widehat{W}(0,0,\alpha,1) =0 \ . 
\ee
For each solution the Nullstellensatz then gives us a factorisation
$\widehat{W}=L_1F_1 + L_2F_2+L_3F_3$. At the Gepner point, the
solutions for $\alpha$ are given by the $(k_3+2)$th roots of $-1$ and
the factorisation reduces to the transposition brane, as
required. 

Of particular interest is the case where $w_3=w_4\neq 1$. In this case we
are geometrically in the situation that the projective space acquires
a singularity and the hypersurface intersects with it for generic
complex structure deformations. The intersection points are then exactly
given by $L_1=L_2=L_3=0$, where $\alpha$ solves (\ref{alpha}).

We should note that in both cases, the factorisations that can be
deformed do not couple to the charges in the untwisted sector (that
are in turn in one-to-one correspondence to the polynomial deformation
moduli). We shall see later on that this is indeed a necessary
condition for the deformation to be possible.

\subsubsection{Enhancement of the Picard lattice}

As we have seen in section \ref{geom}, the rank $\rho$ of the Picard 
lattice is enhanced for hypersurfaces in weighted projective space. We
would now like to understand this enhancement from the point of view
of matrix factorisations. 

Let us first discuss the part of the charge lattice that is inherited 
from the embedding space. For this, we consider the tensor product 
factorisations that correspond to the RS-branes. As we have just seen,
these factorisations exist for arbitrary deformations of the
superpotential. We expect on general grounds 
\cite{DougDia,T,Mayr,GJ} that these branes carry precisely all the
charges that can be obtained as pullbacks from the embedding space. If
this is so, then it follows from the discussion in section \ref{geom}
that their rank should be  
\be
{\rm rk(tensor)}= 3+ \sum_{i<j} 
\Bigl( {\rm gcd}(w_i,w_j) -1 \Bigr)\ .
\ee
Here, the $3$ represents the D0, D4 and generic D2 charge, and the
other contribution comes from the resolution of the singularities of
the embedding weighted projective space. We have verified that this
relation is indeed true for all $14$ examples; the relevant rank is
given in the penultimate column of table~1. This gives good support
to the assertion that the tensor factorisations account precisely for
the charges that can be understood in terms of the embedding
projective space.
\smallskip

As we have seen, a $\ZZ_m$ singularity of the embedding space can lead  
to an enhancement of the rank of the Picard lattice of the
hypersurface by a multiple of $m-1$ if the hypersurface intersects the
exceptional locus in more than one point. For example, the rank of the
Picard lattice of the example $\IP_{1,1,2,2}[6]$ was shown to have
$4$ as a lower bound, where $3$ holomorphic curves come from
replacing the points $z_1=z_2=0$, $z_3-\eta z_4=0$, $\eta^3=-1$ by
$\IP_1$'s. It is now natural to believe that these 
additional charges can be obtained as matrix factorisations of 
type (34), where $\eta$ appears as the parameter in the (34) 
part of the factorisation.  

In order to check this claim we have verified that for each $\eta$,
the rank of the charge lattice of the tensor and (34)
factorisations is bigger by one than that of the tensor
factorisations. Furthermore, if we consider two different 
(34) factorisations with different $\eta$, the rank is increased
by $2$, but considering all three different values does not
increase the rank any further (since the symmetric combination of the
three $\eta$ values is already part of the tensor
charges). Furthermore, as we have just seen, all of these
factorisations can be defined for arbitrary complex structure 
deformations. This explains from a matrix factorisation point of view
that for $\IP_{1,1,2,2}[6]$ $\rho\geq 4$ at a generic point in the
complex structure moduli space. 

We have studied these phenomena also for the other examples. The 
rank of the charge lattice spanned by the (34) branes, where
lcm$(w_3,w_4)=m$ is (for fixed value of $\eta$) always by $m-1$ bigger
than the rank of the tensor product lattice. Furthermore, including
all values of $\eta$ we obtain the generic part of the Picard lattice
that arises because the K3 surface is embedded in the weighted
projective space under consideration. The rank of the charge lattice
that is generated by these factorisations is given in the last column
of table~1; this agrees always with what is expected based on the
geometric analysis of section~2.

\subsection{An infinitesimal analysis}

In section \ref{deformations} we considered special factorisations
that could be globally deformed. We would now like to investigate
under which conditions a given factorisation can at least be
infinitesimally deformed. Given $Q_0$, we want to find a $Q(\psi)$
with $Q(\psi)\rightarrow Q_0$ for $\psi\rightarrow 0$ such  that  
\be
Q(\psi)^2=W+\psi V
\ee
at least for small $\psi$. We make the analytic ansatz
\cite{Hori:2004} 
\begin{equation}
Q(\psi)=\sum_n\psi^nQ_n\ , \label{ana}
\end{equation}
and obtain to first order
\begin{equation}
\{Q_0,Q_1\}=\psi V \label{ex}\ .
\end{equation}
As $V\cdot\mathbf{1}$ is $Q_0$-closed, this reduces to a cohomology
problem: if $V$ is not exact, then $Q_0$ is obstructed and cannot be
continued. At higher order we obtain similar conditions: since 
\begin{equation}
\{Q_0,Q_n\}=-\sum_{k=1}^{n-1}Q_kQ_{n-k}\label{QH4}
\end{equation}
the right hand side must be $Q_0$-closed as well. In principle,
obstructions may occur at higher orders too, but we have not 
found any examples where higher order obstructions were important.

The above ansatz (\ref{ana}) implies that the deformation is analytic,
but it is conceivable that non-analytic deformations could exist. 
In particular, holomorphicity implies that there is only one smooth
family of brane deformations, but physically there are certainly
situations where more than a single deformation could compensate for
a given bulk perturbation. In such cases we would expect non-analytic
behavior of $Q(\psi)$. We can make a more general ansatz by making
$\psi$ an analytic function of a parameter $\phi$: 
\begin{equation}
\left(\sum_n\phi^nQ_n\right)^2=W+\psi(\phi)V=W
+\sum_nc_n\phi^n\cdot V \ . \label{ana2} 
\end{equation}
However, this new ansatz is in fact only more general than 
(\ref{ana}) if the cohomology of $Q_0$ is non-trivial. (Physically, this 
corresponds to $Q_0$ having fermions in its self-spectrum.)  

While we cannot solve the obstruction problem in general, we can at
least give a necessary condition for the analytic deformation to be
unobstructed.

\subsubsection{A necessary condition for unobstructed
deformations}\label{ss:thm} 

We can make a general statement about the conditions that allow a
brane to be continued: if $Q$ can be continued analytically 
under the deformation $V$, then $Q$ is not charged with respect to the
corresponding RR ground state $\phi$. 

To prove this we start out with (\ref{KL}) in the untwisted
sector. First of all, it is clear that if $r$ is odd, the charge is
always zero. We can thus assume that there is an even number of
factors (as is always the case for the K3 examples). The charge is
given by  
\begin{eqnarray}
{\rm ch}(Q) & = & \oint dx_1\ldots dx_{r}\frac{V\mathrm{Str}
[\partial_1 Q\ldots\partial_{r}Q]}{\partial_1W\ldots\partial_{r}W}
=\oint dx_1\ldots dx_{r}\frac{\mathrm{Str}[V\mathbf{1}
\partial_1 Q\ldots\partial_{r}Q]}{\partial_1W\ldots\partial_{r}W} 
\nonumber \\
& = & \oint dx_1\ldots dx_{r}\frac{\mathrm{Str}[\{Q,A\}\partial_1
  Q\ldots\partial_{r}Q]}{\partial_1W\ldots\partial_{r}W}\ ,
\end{eqnarray}
where we have used that $V$ must be exact if $Q$ can be continued. 
Consider now terms of the form $Q\partial Q$. Since $Q\partial
Q=\partial(Q^2)-\partial Q Q$ and $Q^2=W\mathbf{1}$, 
$$
\oint dx_1\ldots dx_{r}\frac{\mathrm{Str}[\partial_1
Q\ldots\partial_k(Q^2)\ldots\partial_{r}Q]}
{\partial_1W\ldots\partial_{r}W} 
=\oint dx_1\ldots dx_{r}\frac{\partial_i W\mathrm{Str}
[\partial_1 Q\ldots\ldots\partial_{r}Q]}
{\partial_1W\ldots\partial_iW\ldots\partial_{r}W}
$$
and $\partial_iW$ cancels. At all Gepner points,
$W=x_1^{h_1}+\ldots+x_r^{h_r}$, so $x_i$ only appears in the
numerator. The residue integral $\oint dx_i$ it thus zero. (This
argument works also if $W$ is not of the particular form given above,
see \cite{Kapustin:2003}.) This calculation shows that $\partial Q$
and $Q$ anticommute in the supertrace. Pulling $Q$ through all the
factors and using (anti-)cyclicity of the supertrace, we see that $QA$ 
cancels out with $AQ$ and ${\rm ch}(Q)$ is thus zero. 

This proof works for the twisted sector as well. In this case it
suffices to realise that $Q_n$ commutes with $\gamma_M^n$. This
follows from the fact that 
\be
\gamma_M^n \, Q_n(\omega^{nw_i}x_i)=Q_n(x_i)\, \gamma_M^n\ .
\ee
But according to the definition of $Q_n$, only those $x_i$ appear for
which $\omega^{nw_i}=1$. It is also clear that if we insert any
fermionic boundary operator $F$ such that $\{Q,F\}=0$, the charge
remains zero. If there is an odd number of factors now, this is
trivial, otherwise $F$ just provides the additional sign that makes
the trace disappear. 

We note in passing that the globally deformed branes that we discussed
in section 3.3.1 do indeed satisfy this condition.

\subsubsection{A counterexample}\label{ss:thm2}

In the previous subsection we have seen that a necessary condition for
the brane $Q$ not to be obstructed under the deformation $V$ is that
$Q$ is not charged under $V$. One may wonder whether this condition
is also sufficient, but this is not true. As an explicit
counterexample consider the deformation of the (12)(34) brane of the
quartic under the deformation  $x_1^2x_3^2$. If we choose the two
values of $\eta$  to be $\eta_1$ and $\eta_2$, 
$Q^{(1)}=Q_{\eta_1}(x_1,x_2)\odot Q_{\eta_2}(x_3,x_4)$, then their
charge is 
\be\label{ch1}
{\rm ch}(Q^{(1)})(x_1^2x_3^2)=\frac{\eta_1^3\eta_2^3}{16}\ .
\ee
Now define $Q^{(2)}=Q_{\eta_1}(x_1,x_2)\odot Q_{-\eta_2}(x_3,x_4)$,
and consider the superposition $Q$ of these two factorisations 
\begin{equation}
Q=\left(\begin{array}{cccc}0&0&J_1&0\\0&0&0&J_2\\
E_1&0&0&0\\0&E_2&0&0\end{array}\right)\ .
\end{equation}
Because of (\ref{ch1}), this factorisation is then uncharged under
$x_1^2 x_3^2$. Nevertheless it cannot be analytically deformed. If it
could, we would have to find a matrix 
\begin{equation}
X=\left(\begin{array}{cccc}0&0&A&B\\
0&0&C&D\\E&F&0&0\\G&H&0&0\end{array}\right)
\end{equation}
consisting of polynomial block matrices $A,B,\ldots, H$ such that
\be
\{Q,X\}=x_1^2x_3^2\, \mathbf{1}\ .
\ee
This yields eight (matrix) equations. The first and the fifth one are
\begin{eqnarray}
J_1E+AE_1 & =& x_1^2x_3^2\mathbf{1} \nonumber\\
E_1A+EJ_1 & =& x_1^2x_3^2\mathbf{1}\ .\label{nogo}
\end{eqnarray}
These are, however, the very equations we find if we want to deform  
$Q^{(1)}$ itself. On the other hand, we know that $Q^{(1)}$ is charged
under $x_1^2x_3^2$ and thus not analytically deformable, so 
(\ref{nogo}) has no solution. This shows that $Q$ is not analytically
deformable either.

\subsection{Quartics on the orbifold line}

As an interesting application of the above techniques we now want to 
study the line of `very attractive' quartics (\ref{veryattractive})
\cite{Wend05} from a matrix factorisation perspective. First we need
to collect some information regarding the RR charges.

It is easy to see that there are $21$ monomial deformations of the 
quartic superpotential that have integer $U(1)$ charge in the closed
string theory. These correspond to $21$ RR ground
states in the untwisted sector. $19$ of these monomials are of charge
$1$ --- the corresponding RR ground states have charge $0$ and can
couple to both A-type and B-type branes. The remaining two integer
charge monomials $1$ and $x_1^2x_2^2x_3^2x_4^2$ correspond to RR
ground states which have $q_L=q_R\neq0$ and hence can couple only to
A-type branes. Furthermore, each of the twisted sectors gives rise to 
one RR ground state each that can couple to B-type branes. 

Our first task is to find a set of matrix factorisations that span the
full B-type charge lattice. At the Gepner point, such a set is given
by the double transposition branes (12)(34), (13)(24), (14)(23), which
span a charge lattice of rank 22, accounting for the D0, D4 and $20$
D2 branes. This in particular verifies that the Picard lattice at the 
Gepner point has maximal rank (namely 20). In the following, we want
to analyse the deformations of these factorisations along the orbifold
line.

\subsubsection{Deformations of the Gepner point}

It follows from our general discussion above that the D0-brane
factorisation (34) and the tensor factorisation can be extended
over the full complex structure moduli space, and therefore in
particular, also along the orbifold line. These branes only couple to
the three twisted RR charges, and thus account for the three generic
RR-charges (that correspond to the D0, the D4, and the one D2-brane).

Next we observe that the (12)(34) factorisations can also be
continued to arbitrary points on the orbifold line. To see this, we
make the ansatz 
\be
L_1=x_2-\alpha_1 x_1\ , \qquad L_2=x_3-\alpha_2 x_4 \ ,
\ee
and insert $L_1=L_2=0$ into the superpotential.
To obtain a factorisation $W=L_1F_1+L_2F_2$ from this ansatz, we require
that the superpotential vanishes on the locus $L_1=L_2=0$. In the case
at hand this yields the following condition on the parameters
$$
1+\alpha_1^4+a\alpha_1^2=0\ , \qquad 1+\alpha_2^4+b \alpha_2^2=0\ ,
$$
which is solvable for $\alpha_i$ for any value of $a,b$. In
particular, this means that we have found a deformed (12)(34)
factorisation for any value of $a,b$. 

By the same argument we also see that the (13)(24) branes with 
\be
x_1-\eta_1 x_3=0\ , \qquad x_2-\eta_2 x_4=0 \label{apmb}
\ee
can be extended to those $a$ and $b$ that satisfy 
$a \eta_1^2 \eta_2^2 +b=0$. For general parameters $a,b$ however,
(13)(24) and (14)(23) are obstructed, as follows immediately from the
fact that they are charged under the corresponding deformations.  
On the other hand, it is possible to construct a
factorisation of the deformed superpotential by writing it as 
\begin{equation}
W(x)=\underbrace{\left(x_1^2+\frac{a}{2}x_2^2\right)^2+
\left(x_3^2+\frac{b}{2}x_4^2\right)^2}_{h=2}
+\underbrace{\left(1-\frac{a^2}{4}\right)x_2^4
+\left(1-\frac{b^2}{4}\right)x_4^4}_{h=4}\ .
\end{equation}
We can then consider the tensor product of the permutation
factorisation of the first two and the last two terms. For
$a\rightarrow0, b\rightarrow0$, these factorisations reduce to tensor
products of permutation factorisations in $x_1^2, x_3^2$ and $x_2,x_4$ 
respectively. We can similarly combine $x_1^2, x_4^2$ and $x_2,x_3$,
{\it etc.}; there are four different constructions of this type, and 
each accounts for two different charges. Together with the (12)(34)
and the RS branes, they then span a charge lattice of rank $20$.    

Thus we have found a set of factorisations that can be deformed along
the whole orbifold line and whose charges generate a sublattice of 
rank $20$. As follows from the analysis of section \ref{ss:thm}, all
of these factorisations are uncharged under  $x_1^2x_2^2$ and  
$x_3^2 x_4^2$, as is indeed also readily verified. The lattice of
B-type D-branes that are uncharged under these two charges has rank
$20$, and thus the above constructions account already for all of
it. 

On the other hand, it also follows from our analysis of section
\ref{ss:thm}, that {\it any} factorisation that is charged with
respect to $x_1^2x_2^2$ or $x_3^2 x_4^2$ cannot be analytically
deformed. Furthermore, the (13)(24) brane for example
does not have a fermion in its self-spectrum, and thus also 
the non-analytic solution of the type (\ref{ana2}) cannot exist. This 
seems to predict that the corresponding D-brane in the orbifold theory
should also be obstructed. We shall explain in detail in the next
section that this is indeed so.

\section{The conformal field theory description} \label{CFT}
\setcounter{equation}{0}

In order to understand in more detail how the geometric and the matrix
factorisation point of view fit together it is useful to study the
quartic K3 surface (and its orbifold line) directly in conformal field
theory. The conformal field theory we are interested in has two
equivalent descriptions: it can be described as the Gepner model
corresponding to the four-fold tensor product of four $N=2$ minimal
models with $k=2$; on the other hand, the theory is also equivalent to
the $\Zop_4$ orbifold of a $T^4$-torus. The equivalence involves in fact
mirror symmetry. In the following we shall first explain briefly the
relevant Gepner model construction, and then describe in more detail
the torus orbifold realisation and the correspondence between the two
descriptions. Finally we shall describe some of the D-branes from both
points of view, and explain how they deform under the Kaehler
deformations of the orbifold theory.

\subsection{The Gepner model}

The Gepner description is standard \cite{Gepner:1987qi}, so we shall
be fairly brief in the following. (A more
comprehensive introduction to Gepner models can be found in
\cite{Gepner:1989gr}; our conventions are explained in more detail for
example in \cite{Recknagel:1998sb,Brunner:2005fv}.)

The Gepner model of interest is the $\Zop_4$-orbifold of
the four-fold tensor product of $k=2$ minimal models (each having
$c=3/2$, so that the total central charge is $c_{tot}=6$). As usual we
label the representations of the bosonic subalgebra of the $N=2$
superconformal algebra by triples $(l,m,s)$ of integers, where $l$
takes the values $l=0,1,2$, and $m$ and $s$ are defined modulo $8$
and $4$, respectively. The three integers have to obey 
$l+m+s=0\mod 2$. Furthermore there is an identification 
\begin{equation}
(l,m,s)\sim (2-l,m+4,s+2) \ .
\end{equation}
The conformal weight $h$ and the U(1)-charge $q$ of the highest
weight state in the representation $(l,m,s)$ are given by
\begin{equation}\label{weightcharge}
h (l,m,s)= \frac{l (l+2)-m^{2}}{16} +\frac{s^{2}}{8} \mod \mathbb{Z}
, \qquad q (l,m,s)= \frac{s}{2}-\frac{m}{4}  \mod
2\mathbb{Z} .
\end{equation}
Representations with $s$ even belong to the Neveu-Schwarz sector,
while those with $s$ odd belong to the Ramond sector. 

\noindent The space of states of the full theory is of the form 
\begin{equation}
\bigotimes_{i=1}^{4} 
\mathcal{H}_{(l_{i},m_{i}+n,s_{i})}\otimes
\bar{\mathcal{H}}_{(l_{i},m_{i}-n,\bar{s}_{i})} \ ,
\end{equation}
where $n=0,1,2,3$ denotes the twisted sector, and
$s_{i}$ (and $\bar{s}_{i}$) are all either even (NS) or all odd
(R). The labels $m_{i}$ are subject to the integrality condition    
\begin{equation}\label{intcond}
\sum_{i=1}^{4} \frac{m_{i}}{4} \in \mathbb{Z}\ .
\end{equation}
Finally, we may impose the type 0B GSO-projection which requires that 
\begin{equation}\label{GSO}
\sum_{i=1}^{4} \Big( \frac{s_{i}}{2}+\frac{\bar{s}_{i}}{2}\Big) \in  
2\mathbb{Z} \ .
\end{equation}

Of particular importance are the RR ground states of this
theory. Ramond ground states are characterised by the property that
their conformal weight $h$ equals $c/24$. One can easily show that the
ground state of the sector $(l,m,s)$ is a Ramond ground state if it is
of the form $(l,l+1,1)$ or $(l,-l-1,-1)$. The above Gepner model
possesses $24$ RR ground states; in each of the three twisted sectors
($n=1,2,3$) there is one RR ground state which is the state 
\begin{equation}
(n-1,n,1)^{\otimes 4} \otimes \overline{(n-1,-n,-1)}^{\otimes 4} \ . 
\end{equation}
The remaining $21$ RR ground states come from the untwisted $n=0$
sector; if we associate to the R ground state representations
\begin{equation}
(0,1,1) \leftrightarrow 1 \ , \qquad 
(1,2,1)_i \leftrightarrow x_i \ , \qquad
(2,3,1)_i \leftrightarrow x_i^2 \ ,\label{RRmono}
\end{equation}
where the index $i$ refers to the $i$th factor, then we have the state
$1$, $x_1^2 x_2^2 x_3^3 x_4^2$, as well as the $19$ monomials in $x_i$
that are of degree $4$.

\subsection{The torus orbifold}

The torus in question is simply the orthogonal product of four
circles, which initially all have the self-dual radius and vanishing
$B$-field. For the following it is convenient to write this $4$-torus
as  $T^4=T^2\times T^2$. The $\Zop_4$ orbifold acts by a
counterclockwise rotation by $90$ degrees in the first $T^2$, and by a
clockwise rotation by $90$ degrees in the second. We denote the four
real directions by $y^i$ with $i=1,2,3,4$, and introduce complex 
coordinates in the usual way: $z^1 = y^1 + i y^2$ and 
$z^2 = y^3 + i y^4$. The $\Zop_4$ action is then 
\be \label{z4}
g: z^1 \mapsto e^{\frac{2\pi i}{4}} \, z^1 \ , \qquad 
z^2 \mapsto e^{- \frac{2\pi i}{4}} \, z^2  \ .
\ee
We denote the fermionic fields by $\chi^i$ and $\tilde\chi^i$, where
$i=1,2,3,4$. The corresponding complex fields are then
\be\label{comfer}
\psi^1 = \frac{1}{\sqrt{2}} \left( \chi^1 + i \chi^2 \right) 
\quad
\bar\psi^1 = \frac{1}{\sqrt{2}} \left(\chi^1 - i \chi^2\right)
\quad
\psi^2 = \frac{1}{\sqrt{2}} \left( \chi^3 + i \chi^4 \right)
\quad
\bar\psi^2 = \frac{1}{\sqrt{2}} \left( \chi^3 - i \chi^4 \right) 
\ee
with similar formulae for the right-moving fields, $\tilde\psi^1$,
{\it etc.}

The momentum ground states of the torus are labelled by four momentum
numbers $n_i$, $i=1,2,3,4$, and four winding numbers $w_j$,
$j=1,2,3,4$. For the $i$th direction the left- and right-moving
momenta are then 
\be
\left( p_L^i,p_R^i \right) = 
\left( \frac{n_i}{2R_i} + w_i R_i, \frac{n_i}{2R_i} - w_i R_i \right)
\ ,
\ee
where initially all $R_i=\frac{1}{\sqrt{2}}$ in our conventions.  
On the ground states, the $\Zop_4$ action maps 
\be\label{modeaction}
(n_1,w_1,n_2,w_2,n_3,w_3,n_4,w_4) \mapsto 
(-n_2,-w_2,n_1,w_1,n_4,w_4,-n_3,-w_3) \ .
\ee
This symmetry requires only that $R_1=R_2$ and $R_3=R_4$, but neither
needs to take the self-dual value. Thus there is a two
(real)-dimensional space of deformations that preserve the orbifold
symmetry. One should expect on general grounds that this is only part
of a two (complex)-dimensional space of deformations, and this is
indeed so. One easily sees that one can also switch on an arbitrary
$B$-field in either of the two $T^2$: if we concentrate on the
first $T^2$, then the momenta are of the form
\begin{eqnarray}
\left(p_L^1,p_L^2 | p_R^1,p_R^2 \right)  
& = & \left( \frac{n_1}{2R} + w_1 R + B w_2 , 
\frac{n_2}{2R} + w_2 R - B w_1 \right|  \nonumber \\
&& \qquad \qquad \qquad \left. 
\frac{n_1}{2R} - w_1 R + B w_2, 
\frac{n_2}{2R} - w_2 R - B w_1 \right)\ .
\end{eqnarray}
It is then easy to see that the spectrum is invariant under the 
$\Zop_4$-action 
\be
(p_L^1,p_R^1) \mapsto (p_L^2,p_R^2) \mapsto (-p_L^1,-p_R^1) 
\mapsto (-p_L^2,-p_R^2) \ .
\ee
In fact, this action still corresponds precisely to the action (on the
first two coordinates) of (\ref{modeaction}). 

In the following we shall mainly concentrate on the theory where the 
$B$-field vanishes and all the radii take the self-dual value
$R_i=\frac{1}{\sqrt{2}}$; this is the theory that corresponds
precisely to the Gepner model $(2)^4$. Unless mentioned otherwise this
is what we shall call the torus orbifold in the following.

\subsection{A partial dictionary}

Before proceeding we shall match a few low-lying states in order to
understand how the identification works. In the untwisted NS-NS sector
of the torus orbifold, the lowest lying states is the vacuum with
$h=\bar{h}=0$, as well as four states of $h=\bar{h}=1/4$. The latter
are the $\Zop_4$-orbits of the states for which the only non-vanishing
momentum and winding number is $n_1=1$, or $w_1=1$ or $n_3=1$ or 
$w_3=1$. In the first ($g$) and third ($g^3$) twisted sector of the
torus orbifold, there are $4$ fixed points each, and each of them has
(in the NS-NS sector) ground state energy $h=\bar{h}=1/4$. (There are
$10$ $\Zop_4$-orbifold invariant $\Zop_2$-fixed points in the second 
($g^2$) twisted sector, but their ground state energy is higher.) In
total the torus orbifold therefore has $12$ NS-NS states with
$h=\bar{h}=1/4$.  

In the $N=2$ orbifold (the Gepner model), all of these states appear
in the untwisted ($n=0$) sector; the vacuum obviously corresponds to
the ground state of the trivial representations, and the states with 
$h=\bar{h}=1/4$ are the ground states in the sectors
\begin{equation}\label{h=1/4}
(1,\pm 1,0) \otimes (1,\mp 1, 0)\otimes (0,0,0) \otimes (0,0,0)\ ,
\end{equation}
where the two non-trivial representations may appear in any two of the
four factors (there are six different possibilities), and the two
signs are correlated. [In the above we have only written the
left-moving representations; since $n=0$ the right-moving
representations are simply equal.] 

\subsubsection{RR ground states}

It is also instructive to understand how the $24$ RR ground states of
the $N=2$ Gepner model that were described at the end of the previous
section appear in the torus orbifold. In the untwisted sector of the
torus orbifold we have eight fermionic zero modes, namely $\chi^i_0$
and $\tilde\chi^i_0$, or the corresponding complex modes defined by 
(\ref{comfer}). We combine them into creation and annihilation
operators by defining  
\begin{eqnarray}
\psi_1^\pm & = & \frac{1}{\sqrt{2}} 
\left( \psi^1_0 \pm i \tilde\psi^1_0 \right)  = 
\half \left(\chi^1_0 + i \chi^2_0 \right) \pm i 
\half \left(\tilde\chi^1_0 + i \tilde\chi^2_0 \right) \nonumber \\
\bar\psi_1^\pm & = & \frac{1}{\sqrt{2}} 
\left( \bar\psi^1_0 \pm i \tilde{\bar{\psi}}^1_0 \right)  = 
\half \left(\chi^1_0 - i \chi^2_0 \right) \pm i 	
\half \left(\tilde\chi^1_0 - i \tilde\chi^2_0 \right) \nonumber \\
\psi_2^\pm & = & \frac{1}{\sqrt{2}} 
\left( \psi^2_0 \pm i \tilde\psi^2_0 \right)  = 
\half \left(\chi^3_0 + i \chi^4_0 \right) \pm i 
\half \left(\tilde\chi^3_0 + i \tilde\chi^4_0 \right) \nonumber \\
\bar\psi_2^\pm & = & \frac{1}{\sqrt{2}} 
\left( \bar\psi^2_0 \pm i \tilde{\bar{\psi}}^2_0 \right) =
\half \left(\chi^3_0 - i \chi^4_0 \right) \pm i 
\half \left(\tilde\chi^3_0 - i \tilde\chi^4_0 \right) \ . \nonumber  
\end{eqnarray}
We define $|0\rangle_{RR}$ to be the state that is annihilated by the
$-$ modes, {\it i.e.}
\begin{equation}
\psi_j^- |0\rangle_{RR} = \bar\psi_j^- |0\rangle_{RR} = 0 \ , \qquad 
j=1,2 \ .
\end{equation}
The space of RR ground states is thus generated by the action of the
$+$-modes from this state. Since there are four creation operators,
the space of RR ground states (before orbifold projection) is 
$16$-dimensional. 

The state $|0\rangle_{RR}$ can be taken to be invariant under the
orbifold action, while the $\psi$-modes transform as 
\begin{eqnarray}
g\, \psi_1^\pm g^{-1} = e^{\frac{2\pi i}{4}}\, \psi_1^\pm \ ,
&  \qquad & 
g \, \bar\psi_1^\pm g^{-1} = e^{- \frac{2\pi i}{4}}\, \bar \psi_1^\pm
\ , \nonumber \\ 
g \, \psi_2^\pm g^{-1} = e^{- \frac{2\pi i}{4}} \, \psi_2^\pm \ , 
& \qquad & 
g \, \bar\psi_2^\pm g^{-1} = e^{\frac{2\pi i}{4}} \, \bar\psi_2^\pm 
\ . 
\end{eqnarray}
Of the $16$ RR ground states, there are therefore six states that are
invariant under the $\Zop_4$ orbifold action: in addition to 
$|0\rangle_{RR}$ and 
$\psi_1^+ \bar\psi_1^+ \psi_2^+ \bar\psi_2^+ |0\rangle_{RR}$ they are 
\begin{equation}
\psi_1^+ \bar\psi_1^+ |0\rangle_{RR} \ , \qquad 
\psi_1^+ \psi_2^+ |0\rangle_{RR} \ , \qquad
\bar\psi_1^+ \bar\psi_2^+ |0\rangle_{RR} \ , \qquad 
\psi_2^+ \bar\psi_2^+ |0\rangle_{RR} \ .
\end{equation}
Two linear combinations of these six states can only couple to B-type 
branes, two can only couple to A-type branes, while the remaining two
can couple to either. (A more explicit analysis of the $N=2$ charges
of these states is spelled out in appendix~A.2.) The relevant A-type
branes are the mirror of the D0- and the D4-brane, as well as of two
B-type D2-branes that contribute to the Picard lattice. 

The remaining $18$ RR charges arise from the twisted sector. In order
to describe the twisted sector states it is useful 
to think of the $\Zop_4$ orbifold in two steps as the $\Zop_2$
orbifold of a $\Zop_2$ orbifold. The first $\Zop_2$ orbifold inverts
all four torus coordinates, while the second $\Zop_2'$ orbifold acts
as a rotation by $90$ degrees in the two tori (clockwise in the first,
and anti-clockwise in the second). The first $\Zop_2$ orbifold has, as
usual, $16$ fixed points at $\half(y_1,y_2,y_3,y_4)$, where each $y_i$
is either $0$ or $1$. Only four of these fixed points are invariant 
under the full $\ZZ_4$ orbifold action, namely $(0,0,0,0)$, $(1,1,0,0)$,
$(0,0,1,1)$ and $(1,1,1,1)$. Thus we have four RR ground states in
each of the $g$, $g^2$ and $g^3$-twisted sectors, giving together $12$
RR ground states. This also fits together with the geometric
description of the orbifold: at each $\ZZ_4$ singular point the
geometry looks locally like $\BC_2/\ZZ_4$, which is a singularity of
type $A_3$. Its resolution introduces $3$ exceptional divisors whose
intersection pattern is determined by the corresponding Cartan
matrix. 

The other 12 $\Zop_2$-fixed points form orbits of length $2$ under the
additional $\ZZ_2'$ action, leading to $6$ $\ZZ_2$ fixed 
points of the full $\Zop_4=\Zop_2\times \Zop_2'$ orbifold. (Each of
these introduces a single exceptional divisor.) They therefore only
contribute $6$ states to the $g^2$ 
twisted sector. In total we therefore have $18$ twisted RR 
ground states. All of them correspond to 2-cycles that are part of the
Picard lattice. Together with the two charges that appear in the
untwisted sector we thus see that the rank of the Picard lattice of
the mirror is indeed maximal, $\rho=20$. 

The counting of the RR charges is obviously valid at any point on the 
orbifold line. One might wonder which of the 2-cycles are special, 
so that they cannot be deformed easily in the matrix
factorisation picture. In fact, one would expect that the $18$
2-cycles from the blow-up contribute in a straight-forward manner at
any point along the orbifold line. Indeed, this is in analogy to our 
discussion of hypersurfaces in weighted projective space where the
2-cycles coming from the resolution contribute everywhere in moduli
space. On the other hand, the remaining 2-cycles that come from the
torus do depend more critically on the radii. As we shall see, this
expectation will indeed be borne out.

\subsubsection{Quantum symmetries}

Finally, it is very instructive to identify the quantum symmetries of
the orbifolds on both sides. The quantum symmetry of the torus
orbifold acts on the Gepner model as  
\begin{equation}
e^{\frac{ i \pi}{2} (m_1 + m_2 - s_1 - s_2)} \ .
\end{equation}
In fact, this identification can be read off from the geometric point
of view that will be described below in section~4.4. The states from
the $g^0$ (untwisted) sector of the orbifold theory 
thus correspond to the polynomials 
\begin{equation}
g^0 \longleftrightarrow 1\ ,\quad 
x_1^2 x_2^2\ , \quad 
x_3^2 x_4^2\ , \quad 
x_1^2 x_2^2 x_3^2 x_4^2 \ , 
\end{equation}
and to the $(n=1)$ and $(n=3)$ twisted RR ground states. The four RR
ground states of the $g$-twisted sector are 
\begin{equation}\label{g1}
g^1 \longleftrightarrow 
x_1 x_3 x_4^2\ , \quad 
x_2 x_3 x_4^2\ , \quad 
x_1 x_4 x_3^2\ , \quad 
x_2 x_4 x_3^2\ ,
\end{equation}
while the corresponding statement for the $g^3$-twisted sector is 
\begin{equation}\label{g3}
g^3 \longleftrightarrow 
x_1^2 x_2 x_3\ , \quad 
x_1^2 x_2 x_4\ , \quad
x_2^2 x_1 x_3\ , \quad 
x_2^2 x_1 x_4\ .
\end{equation}
All other RR ground states come from the $g^2$ twisted sector. 
\smallskip

Conversely, we can also identify the quantum symmetry of the $N=2$
orbifold that appears in the construction of the Gepner model, on
the torus side: as will become clear from the detailed analysis of the
appendix~A.2 it seems to be given by the $\Zop_4$ rotation by $90$
degrees that only acts on the first $T^2$, but leaves the second $T^2$
invariant. The three RR ground states in the Gepner model that appear
in the twisted sectors ($n=1,2,3$) then correspond to the three RR
ground states of the torus orbifold that have eigenvalues 
$e^{\frac{2\pi i n}{4}}$ under this $90$ degree rotation. The state
with $n=2$ corresponds to a specific $\Zop_4$-invariant combination 
of $\Zop_2$-fixed points from the $g^2$ sector of the torus orbifold,
while the $n=1$ and $n=3$ states arise from the untwisted sector of
the torus orbifold. In fact one easily sees that 
\be\label{n=1}
\begin{array}{lcl}
(n=1) \qquad 
(0,1,1)^{\otimes 4} \otimes \overline{(0,-1,1)}^{\otimes 4} 
& \longleftrightarrow & \psi_1^+ \psi_2^+ |0\rangle_{RR} \\
(n=3) \qquad
(2,3,1)^{\otimes 4} \otimes \overline{(2,-3,1)}^{\otimes 4} 
& \longleftrightarrow & \bar\psi_1^+ \bar\psi_2^+ |0\rangle_{RR} \ .
\end{array}
\ee
This identification will prove very useful below.

\subsection{The Inose point of view}

The equivalence of the $\ZZ_4$ orbifold line with certain
perturbations of the Gepner quartic can also be understood
\cite{Wend05} as an extension of a purely geometric result due to
Inose \cite{Ino76}. This point of view also ties in  
nicely with the identification of the the D-branes of the two
theories. 

Inose discovered that the K3 surface obtained as a resolution of the
toroidal $\ZZ_2$ orbifold is equivalent to a geometric $\ZZ_2$ orbifold
of the quartic K3 at the Gepner point. As before, the $\ZZ_2$ action on
the torus is given by inversion of all $4$ coordinates, whereas the
$\ZZ_2$ action on the hypersurface acts as 
\begin{equation}
\sigma:(x_1,x_2,x_3,x_4)\to (-x_1,-x_2,x_3,x_4)\ .
\end{equation}
This orbifold action has $8$ fixed points $(x_1,x_2,0,0)$ with
$x_1^4+x_2^4=0$, and $(0,0,x_3,x_4)$ with $x_3^4+x_4^4=0$, introducing
8 exceptional $\IP_1$'s. The lines
\begin{equation}\label{sixteen}
x_1-\eta x_2=0\ , \quad x_3-\eta x_4= 0\ ,
\end{equation}
of which one would expect to have a Landau-Ginzburg description in
terms of the corresponding matrix factorisations (12)(34), are
invariant under the orbifold action. According to Inose, the 16 lines
(\ref{sixteen}) correspond to the 16 $\IP_1$'s required to resolve the
$\ZZ_2$ singularities of $T^4/\ZZ_2$.  

To apply Inose's result to the relation between the toroidal $\ZZ_4$
orbifold and the quartic hypersurface we remember that in conformal
field theory any abelian orbifold possesses a quantum symmetry by
means of which one can undo the orbifold. This quantum symmetry acts
on the twisted sectors by phase multiplication, and if we orbifold the
orbifold theory by this quantum symmetry we reobtain the original
theory. In the case at hand, one would like to divide out the
orbifolded quartic by its quantum symmetry to re-obtain the
quartic. \cite{Wend05} identifies this quantum symmetry on the
toroidal side: it is precisely the $\ZZ_2'$ that enhances the $\ZZ_2$
given by coordinate inversion to the $\ZZ_4$ action (\ref{z4}).  

This can now help us to understand to which twisted sectors (of the
torus theory) the branes of the $(2)^4$ model should couple. 
Let us first note that $\sigma$ acts on the states in the Gepner model
as
\be
\sigma: \otimes (l_i,m_i,s_i) \to (-1)^{l_1+l_2} \otimes (l_i,m_i,s_i)
\ .
\ee
As mentioned before, the lines (\ref{sixteen}) on the quartic
hypersurface are invariant under the $\ZZ_2$ action induced by
$\sigma$. This means that also the corresponding matrix factorisations
are invariant. In conformal field theory language, the boundary states
corresponding to the $(12)(34)$ transposition branes are thus
invariant under the $\sigma$-orbifold operation, and need to be
`resolved' by adding a contribution from the twisted sector, with a
choice of sign reflecting the freedom to pick a representation on the 
Chan-Paton labels. 

Going back to the covering theory by dividing out by the quantum
symmetry, these resolved boundary states then form orbits under the
quantum symmetry and need not be resolved again. On the torus side,
the same should happen, and the corresponding boundary states should
therefore only couple to the $\Zop_2$ fixed points points, but not to
the $g$ or $g^3$ twisted sector of the torus orbifold. In fact, this
is in agreement with the above identification (\ref{g1}) and 
(\ref{g3}) since it follows from (\ref{KL}) that the $(12)(34)$
fctorisations are not charged under these monomials.

On the other hand the (13)(24) matrix factorisations are not invariant
under $\sigma$, and therefore form orbits under $\sigma$-orbifold
action. In turn, they therefore need to be resolved under the quantum
symmetry orbifold. On the torus side, we should therefore expect that
the corresponding boundary states do couple to the $g$ and $g^3$
twisted sectors; again this is in agreement with the 
identifications (\ref{g1}), (\ref{g3}) and the charge formula
(\ref{KL}).

\subsection{Some simple D-branes}\label{ss:CFTsimple}

Having understood at least in parts the dictionary between the Gepner
model and the torus orbifold description, we now want to explain the
torus description of certain classes of branes in Gepner models. In
particular, we want to study the tensor product (RS) branes 
\cite{Recknagel:1998sb} and the permutation branes
\cite{Recknagel:2002qq} whose matrix factorisation description was
explained in
\cite{Ashok:2004zb,Ashok:2004xq,Brunner:2005fv,EGJ,Enger:2005jk}. 
For a related theory, namely $T^2/\Zop_4$, this analysis was recently
performed in \cite{Dell'Aquila:2005jg} (see also \cite{SC}). 

\subsubsection{The tensor branes}

The simplest D-branes of the Gepner model are the RS D-branes that
correspond to tensor products of rank $1$ factorisations of the
separate factors. In analogy to the correspondence between pure D6
branes and their images under the Gepner monodromy and the $L_i=0$ RS
states one expects that these states correspond geometrically to D4
branes wrapped on the quartic hypersurface. Via mirror symmetry, they
are mapped to D2 branes on the torus orbifold. To be more precise,
these D2 should have one Neumann and one Dirichlet direction in each
torus. In fact, one easily sees that the RS-branes do not couple to
the twelve NS-NS states of (\ref{h=1/4}); the RS-branes therefore
cannot be D0- or D4-branes and thus must be D2-branes.

The RS D-branes only couple to the RR ground states
in the twisted sectors ($n\ne 0$). Given the identification 
(\ref{n=1}) as well as the explicit formulae for these boundary
states (see for example \cite{Brunner:2005fv} whose conventions we
employ in the following) we thus know that the RS branes with $L_i=0$
couple to  
\be\label{tensorRR}
|{\rm RS} \rangle\!\rangle^0 \simeq 
\left( e^{-\frac{\pi i \hat{M}}{4}} \, \psi^+_1 \psi^+_2 + 
e^{\frac{\pi i \hat{M}}{4}} \, \bar\psi^+_1 \bar\psi^+_2 \right)
|0\rangle_{RR}  \,.
\ee
Since $\hat{M}$ is even, these D-branes therefore only couple to
differences and sums of these two torus states. With the
identification of the previous section, it is furthermore clear that
these are the only RR ground states of the untwisted sector of the
torus orbifold to which these D-branes couple. [The other such states
arise in the $n=0$ sector of the $N=2$ orbifold, to which the
RS-branes do not couple.] 

In order to determine their orientation we rewrite 
(\ref{tensorRR}) in terms of the real coordinates. We find
\begin{eqnarray}\label{tenglue}
\left(\psi^+_1 \psi^+_2  +  \bar\psi^+_1 \bar\psi^+_2 \right)
|0\rangle_{RR} & = & (\chi_1^+ \chi_3^+ - \chi_2^+ \chi_4^+)
|0\rangle_{RR} \nonumber \\
i \left(\psi^+_1 \psi^+_2  -  \bar\psi^+_1 \bar\psi^+_2 \right)
|0\rangle_{RR} & = & (\chi_2^+ \chi_3^+ + \chi_1^+ \chi_4^+)
|0\rangle_{RR} \,,
\end{eqnarray}
where $\chi_i^+ = \chi_i + i \tilde{\chi}_i$. 
In the former case (which corresponds to $\hat{M}=0$ mod $4$), the
branes are the superposition of branes with Neumann directions along
$y^1$ and $y^3$, and branes with Neumann directions along $y^2$ and
$y^4$. In the latter case the relevant D2-branes have Neumann
directions along $y^2$ and $y^3$, and Neumann directions along $y^1$
and $y^4$. These superpositions are then $\Zop_4$-orbifold invariant.

\subsubsection{The transposition branes}

The next simplest class of D-branes are the transposition branes
corresponding to the permutation $(ij)$, where $i\ne j$. These branes
couple to the same RR ground states as the tensor branes. Their
coupling is however different: taking into account the subtle factor
in the relative overlaps to the tensor branes (see eq.\ (5.7) of 
\cite{Brunner:2005fv}), we find that the branes (again with $L_i=0$)
couple instead of (\ref{tensorRR}) to 
\be\label{tensor12}
|(ij) \rangle\!\rangle^0 \simeq 
\frac{1}{\sqrt{2}} \left( 
e^{-\frac{\pi i (\hat{M}+1)}{4}} \, \psi^+_1 \psi^+_2 + 
e^{\frac{\pi i (\hat{M}+1)}{4}} \, \bar\psi^+_1 \bar\psi^+_2 \right) 
|0\rangle_{RR}  \,.
\ee
Since $\hat{M}$ is even, these branes therefore couple to different
linear combinations; in terms of the real coordinates, the ground
state is proportional to 
\begin{eqnarray}\label{transglue}
\left(\psi^+_1 \psi^+_2 + i \bar\psi^+_1 \bar\psi^+_2 \right) 
|0\rangle_{RR}  & = & 
(1+i) \left[ \chi_1^+ (\chi_3^+ + \chi_4^+) + 
\chi_2^+ (\chi_3^+ - \chi_4^+ ) \right] |0\rangle_{RR}  \nonumber \\
\left(\psi^+_1 \psi^+_2 - i \bar\psi^+_1 \bar\psi^+_2 \right) 
|0\rangle_{RR}  & = & 
(1-i) \left[ \chi_1^+ (\chi_3^+ - \chi_4^+) -
\chi_2^+ (\chi_3^+ + \chi_4^+ ) \right] |0\rangle_{RR}  \ .
\nonumber
\end{eqnarray}
These branes are therefore superpositions of D2-branes that have
Neumann directions along $y^1$ and $y^3\pm y^4$, and branes with Neumann
directions along $y^2$ and $y^3\mp y^4$. 

\subsubsection{The double transposition branes}

The last simple class of branes corresponds to the product of two
transpositions, {\it i.e.} to the permutation $(ij)(kl)$ with
$i,j,k,l$ all mutually distinct. These branes also couple to untwisted
($n=0$) RR ground states, and may therefore also couple to additional
RR ground states of the untwisted ($g^0$) torus orbifold. As regards
the two RR ground states coming from $n=1$ and $n=3$, their coupling
is now (again for $L_i=0$) 
\be\label{(ij)(kl)RR}
|(ij)(kl) \rangle\!\rangle^0 \sim 
\frac{1}{2} 
\left( e^{-\frac{\pi i (\hat{M}+2)}{4}} \, \psi^+_1 \psi^+_2 + 
e^{\frac{\pi i (\hat{M}+2)}{4}} \, \bar\psi^+_1 \bar\psi^+_2 \right)
|0\rangle_{RR}  \,.
\ee
The more detailed interpretation however depends on which permutation
is considered. 
\medskip

\noindent {\bf The case (12)(34):} In this case it follows from the
identification of (\ref{g1}) and (\ref{g3}) that the (12)(34) branes
do {\it not} couple to any RR ground states of the first ($g^1$) or 
third ($g^3$) twisted sector of the orbifold. This implies that they
cannot be $\Zop_4$-fractional branes, and thus that they must
correspond to the superpositions of at least two D2-branes. The
orientation of the two D2-branes is then as described for the tensor
branes in 4.4.1. [The tension of the (12)(34) branes is smaller by a
factor of two than that of the tensor branes; this suggests that the
latter are actually superpositions of four such branes, while the
(12)(34) only involve  two D2-branes.]
\medskip

\noindent {\bf The cases  (13)(24) and (14)(23):} In either case, the 
identification of (\ref{g1}) and (\ref{g3}) now implies that these
branes {\it do} couple to the  first ($g^1$) and third ($g^3$) twisted
sector of the orbifold. Thus they should correspond to `fractional' 
branes. We also know that the $(ij)(kl)$ branes may couple to additional
RR ground states in the untwisted sector of the torus orbifold. In
fact, since the set of all (double) transposition branes account for all
RR charges, at least some of the (13)(24) and (14)(23) branes 
{\it must} couple to these states. From the point of view of the
orbifold description, the relevant RR ground states are identified in 
appendix~A.2. This then suggests that the RR ground states of some of
these boundary states are proportional to  
\begin{eqnarray}
|B1\rangle\!\rangle^0 & \simeq & 
(1 + \psi^+_1 \psi^+_2) \, (1 + \bar\psi^+_1 \bar\psi^+_2)
|0\rangle_{RR} \nonumber  \\
|B2\rangle\!\rangle^0 & \simeq & 
(\psi^+_1 + \bar\psi^+_2) \, (\psi^+_2 + \bar\psi^+_1 )
|0\rangle_{RR}\,. \nonumber 
\end{eqnarray}
These ground states satisfy the gluing conditions 
\begin{equation}\label{1324}
\begin{array}{lll}
(\chi_1 - i \tilde{\chi}_3) |B1\rangle\!\rangle^0 =0 & \qquad & 
(\chi_2 + i \tilde{\chi}_4) |B1\rangle\!\rangle^0 =0 \\[2pt]
(\chi_3 + i \tilde{\chi}_1) |B1\rangle\!\rangle^0 =0 & \qquad & 
(\chi_4 - i \tilde{\chi}_2) |B1\rangle\!\rangle^0 =0 \ ,\\[6pt]
(\chi_1 + i \tilde{\chi}_3) |B2\rangle\!\rangle^0 =0 & \qquad & 
(\chi_2 - i \tilde{\chi}_4) |B2\rangle\!\rangle^0 =0 \\[2pt]
(\chi_3 + i \tilde{\chi}_1) |B2\rangle\!\rangle^0 =0 & \qquad & 
(\chi_4 - i \tilde{\chi}_2) |B2\rangle\!\rangle^0 =0 \ .
\end{array}
\end{equation}
The corresponding D2-branes lie diagonally across the two $T^2$s and
are by  themselves $\Zop_4$-invariant (as should be the case for 
$\Zop_4$-fractional branes!). On the other hand, their area is twice 
that of one of the two D2-branes that appears in the description of
the (12)(34) brane. This is then in accord with the fact that the
tension of the (12)(34) brane agrees with that of the (13)(24) and the
(14)(23) branes.

\subsection{Deforming D-branes}

So far we have (partially) identified the D-branes of the orbifold
theory at the Gepner point with certain classes of matrix
factorisations. In terms of the orbifold theory, it is now not
difficult to describe how these D-branes behave as we vary
the radii or the $B$-fields. 

First of all, it is clear that nothing much of interest happens for
the branes that correspond to the tensor factorisations
(section~4.5.1), the single transposition factorisation
(section~4.5.2), or the (12)(34) branes. In all of
these cases the gluing conditions involve the two $T^2$'s separately,
and the structure of these D-branes is pretty insensitive to changes
of the radii or the $B$-field. 

The situation is however different for the $|B1\rangle\!\rangle$ and 
$|B2\rangle\!\rangle$ branes since they lie diagonally across the two
$T^2$s. As we change the radii of the two $T^2$'s, we generically
change their ratio, which has a significant effect on the
behaviour of these branes. A priori it is not clear how we should
`continue' these D-branes as we vary the closed string parameters, but
there are at least two natural points of views that we can take.  

According to the first point of view, we can simply insist on
preserving the same gluing  conditions (\ref{1324}) (as well as the
corresponding gluing conditions for the bosons) as we vary the radii
and the $B$-fields. By construction, the corresponding D-branes will
then continue to couple to the relevant RR ground states, and will
continue to satisfy the correct $N=2$ gluing conditions. However, as
is well known \cite{DougHull}, the structure of the corresponding
D-branes will depend dramatically on the precise ratio of the radii
and the values of the $B$-fields. Consider for example the 
$|\!|B1\rangle\!\rangle$ brane that is characterised by the 
bosonic gluing conditions corresponding to (\ref{1324}) 
\begin{equation}\label{1324b}
\begin{array}{lcllcl}
(a^1_n -  \tilde{a}^3_{-n}) |\!|B1\rangle\!\rangle & = & 0 \ ,
\qquad \qquad 
(a^2_n +  \tilde{a}^4_{-n}) |\!|B1\rangle\!\rangle & = & 0 \ ,
\\
(a^3_n +  \tilde{a}^1_{-n}) |\!|B1\rangle\!\rangle & = & 0 \ ,
\qquad \qquad 
(a^4_n -  \tilde{a}^2_{-n}) |\!|B1\rangle\!\rangle & = & 0 \ .
\end{array}
\end{equation}
For the original theory for which $R_1=R_2=\frac{1}{\sqrt{2}}$ and
$B_1=B_2=0$, we have for example Ishibashi states on the momentum
ground states for which the only non-vanishing momentum and winding
numbers are $n_3=w_1$ or $w_3=-n_1$ or $n_4=-w_2$ or
$w_4=n_2$. (Obviously the $\Zop_4$-invariant boundary state will
require that we sum over suitable such combinations of Ishibashi
states.)  As we change the radii $R_1$ and $R_2$ or switch on the
$B$-fields, the Ishibashi state whose ground state has only
non-vanishing momentum and winding numbers equal to $n_3=w_1\ne 0$
say, does typically {\it not} satisfy the zero mode part of
(\ref{1324b}) any more. Thus the set of Ishibashi states that
contribute will depend crucially on the closed string parameters. As a
consequence, the same will be true for their tension, {\it etc}. This
fact is also easy to understand geometrically: fixing the gluing
conditions means that we fix the angles with which the D-brane is
oriented in the $13$- and $24$-planes. As we change the ratio of the
radii, the number of times the brane wraps around the torus in the
$13$ and $24$ directions changes erratically.  

In order to avoid this erratic behaviour, we can adopt the second
point of view, namely that we should modify the gluing conditions as
we change the radii or switch on a $B$-field. For example, for
the case when we change the radii $R_1$ and $R_2$, we can consider
\begin{eqnarray}\label{modified}
\left( \begin{array}{cc} a^1_n \\ a^3_n \end{array} \right) 
&= & 
\left( \begin{matrix} \cos 2\theta 
& \sin 2\theta \cr  - \sin 2\theta & \cos 2\theta
 \end{matrix} \right)
\left( \begin{array}{cc} \tilde{a}^1_{-n} \\ 
\tilde{a}^3_{-n} \end{array} \right) \,, \nonumber \\[3pt]
\left( \begin{array}{cc} a^2_n \\ a^4_n \end{array} \right) 
&= & \left( \begin{matrix} \cos 2\theta 
& - \sin 2\theta \cr  \sin 2\theta &  \cos 2\theta
 \end{matrix} \right)
\left( \begin{array}{cc} \tilde{a}^2_{-n} \\ 
\tilde{a}^4_{-n} \end{array} \right) \,,
\end{eqnarray}
where $2 R_1 R_2 = \tan \theta$. It is straightforward to check
that the states with $n_3=w_3\ne 0$, {\it etc.} then satisfy the zero
mode part of (\ref{modified}) for arbitrary values of $\theta$ (not just 
$\theta=\pi/4$). In order to understand the geometric meaning of these
modified gluing conditions, we rewrite them in terms of complex
coordinates. If we define $\alpha^{(1)}_n=a^1_n+i a^2_n$, {\it etc},
we find   
\begin{equation}
\begin{array}{rclrcl}
\alpha^{(1)}_n & = & \cos 2\theta\, \tilde{\alpha}^{(1)}_{-n} 
+ \sin 2\theta\,  \tilde{\bar{\alpha}}^{(2)}_{-n} \qquad \qquad 
\bar\alpha^{(1)}_n & = & \cos 2\theta \,
\tilde{\bar\alpha}^{(1)}_{-n} 
+ \sin 2\theta \, \tilde{{\alpha}}^{(2)}_{-n} \\
\alpha^{(2)}_n & = & \cos 2\theta \, \tilde{\alpha}^{(2)}_{-n}  
- \sin 2\theta \, \tilde{\bar{\alpha}}^{(1)}_{-n} \qquad \qquad
\bar\alpha^{(2)}_n & = &  \cos 2\theta \,
\tilde{\bar\alpha}^{(2)}_{-n}  
- \sin 2\theta  \, \tilde{{\alpha}}^{(1)}_{-n}
\end{array}
\end{equation}
In order to preserve the usual world-sheet $N=1$ algebra, the fermions
have to follow suit, {\it i.e.}
\begin{equation}\label{standard}
\begin{array}{lcrlcl}
\psi^{(1)}_n & = & i\left(\cos 2\theta\, \tilde{\psi}^{(1)}_{-n} 
+ \sin 2\theta \, \tilde{\bar{\psi}}^{(2)}_{-n} \right) \qquad \qquad
\bar\psi^{(1)}_n & = & i \left( \cos 2\theta\, 
\tilde{\bar\psi}^{(1)}_{-n} 
+ \sin 2\theta \, \tilde{{\psi}}^{(2)}_{-n} \right) \\
\psi^{(2)}_n & = & i \left( \cos 2\theta\, \tilde{\psi}^{(2)}_{-n}  
- \sin 2\theta \, \tilde{\bar{\psi}}^{(1)}_{-n} \right) \qquad 
\qquad 
\bar\psi^{(2)}_n & = & i \left( \cos 2\theta\, 
\tilde{\bar\psi}^{(2)}_{-n}  
- \sin 2\theta \, \tilde{{\psi}}^{(1)}_{-n} \right)
\end{array}
\end{equation}
Given the explicit expressions for the $N=2$ and $N=4$ supercharges
of the appendix (see (\ref{N2}) and (\ref{N4})), we can now deduce the
gluing conditions for the supercharges and the $\widehat{su}(2)_1$
currents. Explicitly we find that 
\begin{equation}\label{currentglue}
\left( J^a_n + g \, \tilde{J}^a_{-n} \, g^{-1} \right)
|\!|B1\rangle\!\rangle  = 0 \ , \qquad 
g = \left( \begin{matrix}
\cos 2\theta & - \sin 2\theta \cr \sin 2\theta &
\cos 2\theta 
\end{matrix} \right) \ . 
\end{equation}
[Here we have chosen the convention that the Lie algebra generators
are defined by 
\be
t^+=\left(\begin{matrix} 0 & 1 \cr 0 & 0 \end{matrix}\right)\ , \qquad
t^-=\left(\begin{matrix} 0 & 0 \cr 1 & 0 \end{matrix}\right)\ , \qquad
t^3=\left(\begin{matrix} 1 & 0 \cr 0 & -1 \end{matrix}\right)\ .]
\ee
For all values of $\theta$, the modified boundary state
preserves the $N=4$ superconformal algebra. However, we can ask
whether there is an $N=2$ subalgebra of the $N=4$ algebra for which
the gluing conditions are A-type. In particular, the $U(1)$ current 
$K$ of this $N=2$ subalgebra would then have to satisfy the
gluing condition  
\be
\left( {K}_n - \tilde{K}_{-n} \right) 
|\!|B\rangle\!\rangle  = 0 \ .
\ee
Given (\ref{currentglue}) this means that $K$, regarded as an element
of the Lie algebra of $su(2)$, satisfies
\be
g K g^{-1} = - K \ .
\ee
It is easy to see that such a $K$ only exists if $\cos 2\theta=0$, 
{\it i.e.} for $2R_1R_2=1$.  Thus unless $2R_1 R_2=1$ (for
$B_1=B_2=0$), it is not possible to extend the boundary state 
$|\!|B1\rangle\!\rangle$ in this manner while preserving $N=2$ A-type
supersymmetry. On the other hand, if $2R_1 R_2=1$ (for $B_1=B_2=0$) is
satisfied, one can easily see that the usual $N=2$ subalgebra
continues to satisfy an A-type gluing condition. A similar analysis
also works for other modifications of the gluing conditions, as well
as for $|\!|B2\rangle\!\rangle$. 

These findings reflect now very nicely the results we obtained from
the matrix factorisation point of view. There we saw that for generic 
deformations parametrised by $a$ and $b$ it was not possible to extend
the (13)(24) and (14)(23) factorisations. However, there were special
directions for which an extension was possible: for example, as was
mentioned in section~3.5.1.\ we could extend the (13)(24)
factorisation if $a$ and $b$ satisfy 
\be \label{abrel}
a \eta_1^2 \eta_2^2 + b = 0\ , 
\ee
and a similar condition holds for the (14)(23) factorisations. In
terms of the orbifold theory the condition (\ref{abrel}) means that
the two tori have the same Kaehler parameter (up to an
SL$(2,\Zop)$ transformation). This follows from the fact that the
relation between $a$ and the Kaehler parameter $\rho_1$ of the first
torus is \cite{GS} 
\be
j(\rho_1) = \frac{1}{27 \cdot 4} \frac{(a^2+12)^3}{(a^2-4)^2} \ ,
\ee
with an identical relation between $b$ and the Kaehler parameter 
$\rho_2$ of the second torus. In particular, the two Kaehler
parameters only depend on $a^2$ and $b^2$, respectively, and thus
(\ref{abrel}) implies that  $j(\rho_1)=j(\rho_2)$. This is in
particular the case if $2R_1R_2=1$ (at $B_1=B_2=0$). Thus the special
unobstructed deformations correspond to each other.   

It would be interesting to have a more precise dictionary between the 
different matrix factorisations (including the values of $\eta_1$ and
$\eta_2$, {\it etc.}) and the orbifold D-branes. This would allow one
to check these identifications in even more detail.

\section{Conclusions}

In this paper we have studied B-type D-branes on K3 from three
different points of view: using geometrical methods
(section~2), with the help of the matrix factorisation approach
(section 3), and for the $T^4/\Zop_4$ orbifold line in conformal field
theory (section~4). We have shown that the results we obtained from
these different points of view fit very well together. In
particular, we have been able to understand the  generic rank of the
Picard lattice for K3's that are hypersurfaces in weighted projective
space both from a geometrical point of view and using matrix factorisation 
techniques. For the case of the $T^4/\Zop_4$ orbifold line we have
furthermore managed to identify in some detail the different
matrix factorisation with boundary states in the orbifold conformal
field theory. Furthermore, we could understand from both points of
view why certain D-branes are obstructed against deformations of the
bulk theory.  

More generally, we have found a necessary criterion for when a given
matrix factorisation can be analytically extended under a bulk
deformation: this is only possible if the factorisation is uncharged
under the RR field that corresponds to the bulk deformation. It would
be good to understand this condition directly in conformal field
theory.  
\medskip

Among other things, our results demonstrate convincingly that the
matrix factorisation approach is a very powerful method to study
D-branes at generic points in the moduli space where the traditional
conformal field theory techniques are unavailable. One may hope to be
able to push this further and deduce more global properties about
D-branes on Calabi-Yau manifolds. This should, in particular, be
possible for D-branes on K3 where we have extended supersymmetry.

\section*{Acknowledgements}

This research has been  partially supported by a TH-grant 
from ETH Zurich, the Swiss National Science Foundation and the Marie
Curie network `Constituents, Fundamental Forces and Symmetries of the
Universe' (MRTN-CT-2004-005104). We thank Stefan Fredenhagen, 
Wolfgang Lerche und Cornelius Schmidt-Colinet for useful
discussions. Some parts of this paper are based on the Diploma thesis
of one of us (CAK).



\section*{Appendix}

\appendix

\section{The RR ground states of the torus theory}
\renewcommand{\theequation}{A.\arabic{equation}}
\setcounter{equation}{0}

In this appendix we collect various facts about the RR ground states
of the torus orbifold. In particular, we exhibit the underlying $N=2$ 
and $N=4$ superconformal symmetry in the R sector of this theory. In
the second subsection we explain the dictionary between the torus RR
ground states and the corresponding states in the Gepner model in some
detail.

\subsection{The $N=2$ and $N=4$ algebras}

To fix notation, let us first consider a single $T^2$ with $c=3$. The
left-moving (complex) bosonic and fermionic modes are denoted by
$\alpha_m$, $\bar\alpha_m$, $\psi_n$ and $\bar\psi_n$. The bosonic
modes satisfy the commutation relations 
\begin{equation}
[\alpha_m,\alpha_n] = 0 = [\bar\alpha_m,\bar\alpha_n] \,, \qquad
[\alpha_m,\bar\alpha_n] = m \delta_{m,-n} \,,
\end{equation}
and the fermionic modes the anti-commutation relations 
\begin{equation}
\{\psi_m,\psi_n\} = 0 = \{\bar\psi_m,\bar\psi_n\} \,, \qquad
\{\psi_m,\bar\psi_n\}= \delta_{m,-n} \,.
\end{equation}
The two (chiral) R ground states carry $N=2$ quantum numbers 
$h=1/8$ and $q=\pm 1/2$, and are mapped into one another by the action
of the fermionic zero modes
\begin{eqnarray}
\psi_0  \left|\frac{1}{8},-\frac{1}{2} \right\rangle = 0  & \qquad 
& 
\psi_0  \left|\frac{1}{8},\frac{1}{2} \right\rangle = 
\left|\frac{1}{8},-\frac{1}{2} \right\rangle \nonumber \\
\bar\psi_0  \left|\frac{1}{8},\frac{1}{2} \right\rangle = 0  & \qquad  
& 
\bar\psi_0  \left|\frac{1}{8},-\frac{1}{2} \right\rangle = 
\left|\frac{1}{8},\frac{1}{2} \right\rangle \,.
\end{eqnarray}
Here the $N=2$ generators are defined, in terms of the free bosons and
fermions, as (see for example \cite{Gaberdiel:2004nv})
\be\label{N2}
\begin{array}{lcl}
L_n & = & {\displaystyle \sum_{m} :\alpha_{n-m} \bar\alpha_{m} : 
+ \frac{1}{2} \sum_m (2m-n) : \bar\psi_{n-m} \psi_m : 
+ \frac{1}{8} \, \delta_{n,0}} \vspace*{0.1cm} \\
J_n & = & {\displaystyle \sum_{m} : \bar\psi_{n-m} \psi_{m}: 
- \frac{1}{2} \, \delta_{n,0}}  \vspace*{0.1cm} \\
G^+_n & = & {\displaystyle \sqrt{2} \, \sum_{m} \alpha_{n-m}
  \bar\psi_m}  \vspace*{0.1cm} \\
G^-_n & = & {\displaystyle
\sqrt{2} \, \sum_{m} \bar\alpha_{n-m} \psi_m \ .} 
\end{array}
\ee
One easily checks that they satisfy the correct $N=2$ algebra
with $c=3$, 
\begin{align*}
  \left[L_m, L_n \right] &=(m-n)L_{m+n} + \tfrac{c}{12}(m^3 -  m)
      \delta_{m,-n}  \\ 
  \left[L_m, J_n \right] &=-nJ_{m+n} \\
  \left[L_m, G_n^\pm \right] &=\left( \tfrac{1}{2}m-n \right)
  G_{m+n}^\pm  \\
  \left[J_m, J_n \right] &= \tfrac{c}{3} m\delta_{m,-n}\\
  \left[J_m, G_n^\pm \right] &=\pm G_{m+n}^\pm  \\
  \left\{ G_{m}^+, G_{n}^-\right\} &= 2 L_{m+n} + (m-n) J_{m+n} +
  \tfrac{c}{3}(m^2 - \tfrac{1}{4})\delta_{m, -n} \\
  \left\{G_{m}^+, G_{n}^+ \right\} &=
   \left\{G_{m}^-, G_{n}^-\right\}=0\ .
\end{align*}
Furthermore, they have the correct $N=2$ eigenvalues on the above
states. [Note that the normal ordering for the fermions is defined by 
$:\bar\psi_m \psi_n:=\bar\psi_m \psi_n$ for $m\leq n$ and 
$:\bar\psi_m \psi_n:= - \psi_n \bar\psi_m$ for $m>n$.] It is also
obvious that the above states are annihilated by the zero 
modes $G^\pm_0$, as they must be.

For the case of interest to us, we have two such tori, and thus in
fact an $N=4$ algebra. We denote the relevant free field modes by
$\alpha^{(i)}_n$ and $\psi^{(i)}_n$, where $i=1,2$. The additional
generators of the $N=4$ algebra are the generators $J^\pm_n$ that
enhance the $u(1)$ current $J_n\equiv J^{(1)}_n+J^{(2)}_n$ to an
$\widehat{su}(2)_1$ algebra  
\begin{equation}
J^+_n = \sum_m  :\bar\psi^{(1)}_{n-m} \, \bar\psi^{(2)}_m : \ , \qquad
J^-_n = - \sum_m  :\psi^{(1)}_{n-m} \, \psi^{(2)}_m : \ .
\end{equation}
These generators are obviously orbifold invariant. Note that the
normalisation of the $\widehat{su}(2)_1$ generators is slightly
unusual: they satisfy 
\begin{eqnarray}
{}[J_m,J^\pm_n] & = & \pm \, 2 \, J^\pm_{m+n} \nonumber \\
{}[J^+_m,J^-_n] & = & J_{m+n} + m \, \delta_{m,-n} \nonumber \\
{}[J_m,J_n] & = & 2 \, m \, \delta_{m,-n} \,. \nonumber
\end{eqnarray}
In addition we have two more supercharges ${G'}^\pm_n$ that are
defined by 
\begin{eqnarray}
{G'}^+_{n} & = & \sqrt{2} \, \sum_m  \left( 
\bar\psi^{(1)}_{n-m} \, \bar\alpha^{(2)}_{m} - 
\bar\psi^{(2)}_{n-m} \, \bar\alpha^{(1)}_{m}
\right) \nonumber \\
{G'}^-_{n} & = & \sqrt{2} \, \sum_m  \left( 
\psi^{(1)}_{n-m} \, \alpha^{(2)}_{m} 
- \psi^{(2)}_{n-m} \, \alpha^{(1)}_{m}
\right) \label{N4} \ .
\end{eqnarray}
Together with $G^\pm_n\equiv G^{\pm (1)}_n+G^{\pm (2)}_n$ they then
generate the $N=4$ algebra \cite{Ademollo:1976wv}; in addition to the
above commutation relation of the $N=2$ generators we have
\begin{eqnarray}
\{G^\pm_m,{G'}^\pm_n\} & = & \mp 2 (m-n) J^\pm_{m+n} \qquad \qquad 
\{G^\pm_m,{G'}^\mp_n\} =  0 \nonumber \\
{}[L_m,{G'}^\pm_n] & = & \left(\frac{m}{2} - n \right) {G'}^\pm_{m+n}
\qquad \qquad 
{}[J_m,{G'}^\pm_n]  =  \pm {G'}^\pm_{m+n} \nonumber \\
{}[J^\pm_m,{G}^\pm_n] & = & [J^\pm_m,{G'}^\pm_n] = 0 \nonumber \\
{}[J^\pm_m,{G}^\mp_n] & = & \pm {G'}^\pm_{m+n} \qquad \qquad\qquad 
\quad\, 
{}[J^\pm_m,{G'}^\mp_n]  =  \mp {G}^\pm_{m+n} \nonumber \\
\{{G'}^+_m,{G'}^-_n\} & = & 2 L_{m+n} + (m-n) J_{m+n} + 
2 (m^2 - 1/4) \delta_{m,-n} \ . 
\end{eqnarray}

In the following we shall mostly work with the two $N=2$ algebras
corresponding to the two tori. The RR ground states are then
characterised by their eigenvalues with respect to the two different
U(1)-charges. We shall denote these states by 
$|\pm,\pm\rangle$, where 
\begin{equation}
\begin{array}{lll}
\psi^{(1)}_0 |+,\pm\rangle = |-,\pm\rangle & \qquad & 
\psi^{(1)}_0 |-,\pm\rangle = 0 \\[2pt]
\psi^{(2)}_0 |\pm,+\rangle = \mp |\pm, - \rangle & \qquad & 
\psi^{(2)}_0 |\pm,-\rangle = 0 \\[2pt]
\bar\psi^{(1)}_0 |+,\pm\rangle = 0 & \qquad & 
\bar\psi^{(1)}_0 |-,\pm\rangle = |+,\pm\rangle \\[2pt]
\bar\psi^{(2)}_0 |\pm,+\rangle = 0 & \qquad & 
\bar\psi^{(2)}_0 |\pm,-\rangle = \mp |\pm,+\rangle \,. 
\end{array}
\end{equation}
Note the signs in the second and fourth line --- these are a
consequence of the fact that the fermionic zero modes of the first and
second torus anti-commute.
In the full theory we then have such states for the left- and the
right movers; these will be denoted by 
$|\pm,\pm\rangle \otimes \overline{|\pm,\pm\rangle}$. The action of
the right-moving modes $\tilde\psi^{(j)}_0$ and
$\tilde{\bar\psi}^{(j)}_0$ are given by the same relations as
above. (Note though that the right-moving fermionic modes also
anti-commute with the left-moving fermionic modes, and thus one has to
be careful about relative signs!)

\subsection{The identification with the Gepner model states}

Having set up notation, we can now characterise the RR ground state
$|0\rangle_{RR}$ that is annihilated by the modes 
\begin{equation}
\psi_j^- |0\rangle_{RR} = \bar\psi_j^- |0\rangle_{RR} = 0 \,, \qquad 
j=1,2 \,,
\end{equation}
where
\begin{equation}
\psi_j^\pm = \frac{1}{\sqrt{2}} 
\left( \psi^{(j)}_0 \pm i \tilde{\psi}^{(j)}_0 \right)  \,, 
\qquad
\bar\psi_j^\pm = \frac{1}{\sqrt{2}} \left(
\bar\psi^{(j)}_0 \pm i \tilde{\bar\psi}^{(j)}_0 \right) \,.
\end{equation}
One easily convinces oneself that, in terms of the above basis, 
\begin{equation}
|0\rangle_{RR} = \left( 
|+,+\rangle \otimes \overline{|-,-\rangle} 
- i |+,-\rangle \otimes \overline{|-,+\rangle} 
+ i |-,+\rangle \otimes \overline{|+,-\rangle}
- |-,-\rangle \otimes \overline{|+,+\rangle} \right) \,.
\end{equation}
[In writing this formula we have chosen the convention that the state
with $-$ is bosonic, while that with $+$ is fermionic; so we have for
example 
$$
\tilde\psi^{(1)}_0 |-,-\rangle \otimes \overline{|+,+\rangle} 
= |-,-\rangle \otimes \overline{|-,+\rangle} \,, \qquad
\tilde\psi^{(1)}_0 |-,+\rangle \otimes \overline{|+,+\rangle} 
= - |-,+\rangle \otimes \overline{|-,+\rangle} 
$$
{\it etc}.] Furthermore, we can write the various orbifold invariant
states that are obtained by the action of the $\psi^+$ modes from this
state in terms of the $N=2$ basis. One easily finds
\begin{equation}
\psi^+_1 \psi^+_2 |0\rangle_{RR} = - 2 
|-,-\rangle \otimes \overline{|-,-\rangle} \,, \qquad
\bar\psi^+_1 \bar\psi^+_2 |0\rangle_{RR} = - 2 
|+,+\rangle \otimes \overline{|+,+\rangle} \,.
\end{equation}
These two states therefore have $q=\bar{q}=-1$ and 
$q=\bar{q}=1$, respectively. Since the torus description is related by
mirror symmetry to the Gepner model description, they must correspond
to states in the Gepner model with $q=-\bar{q}=1$ and $q=-\bar{q}=-1$, 
respectively. These are precisely the RR ground states in the twisted
$(n=1)$ and $(n=3)$ sectors, as we had already argued before.
\smallskip

\noindent For the other ground states we obtain 
\small
\begin{eqnarray}
{|0\rangle_{RR}} & = & 
|+,+\rangle \otimes \overline{|-,-\rangle} 
- i |+,-\rangle \otimes \overline{|-,+\rangle} 
+ i |-,+\rangle \otimes \overline{|+,-\rangle}
- |-,-\rangle \otimes \overline{|+,+\rangle} \nonumber \\
{\bar\psi^+_1 \psi^+_1 |0\rangle_{RR}}  & = & 
|+,+\rangle \otimes \overline{|-,-\rangle} 
- i |+,-\rangle \otimes \overline{|-,+\rangle} 
- i |-,+\rangle \otimes \overline{|+,-\rangle}
+ |-,-\rangle \otimes \overline{|+,+\rangle} \nonumber \\
\bar\psi^+_2 \psi^+_2 |0\rangle_{RR} & = & 
|+,+\rangle \otimes \overline{|-,-\rangle} 
+ i |+,-\rangle \otimes \overline{|-,+\rangle} 
+ i |-,+\rangle \otimes \overline{|+,-\rangle}
+ |-,-\rangle \otimes \overline{|+,+\rangle} \nonumber \\
\bar\psi^+_2 \psi^+_2 \bar\psi^+_1 \psi^+_1 |0\rangle_{RR} & = & 
|+,+\rangle \otimes \overline{|-,-\rangle} 
+ i |+,-\rangle \otimes \overline{|-,+\rangle} 
- i |-,+\rangle \otimes \overline{|+,-\rangle}
- |-,-\rangle \otimes \overline{|+,+\rangle} \,. \nonumber 
\end{eqnarray}
\normalsize
The A-type branes of the torus orbifold (that correspond to the B-type
branes of the Gepner model) should therefore couple to the
combinations 
\begin{equation}
|\Psi_1^-\rangle = 
i \left( |-,+\rangle \otimes \overline{|+,-\rangle} - 
|+,-\rangle \otimes \overline{|-,+\rangle} \right) = \frac{1}{2} 
\left( |0\rangle_{RR} - 
\bar\psi^+_2 \psi^+_2 \bar\psi^+_1 \psi^+_1   |0\rangle_{RR}  \right) 
\end{equation}
and
\begin{equation}
|\Psi_2^-\rangle = i \left( |-,+\rangle \otimes \overline{|+,-\rangle} + 
|+,-\rangle \otimes \overline{|-,+\rangle} \right) = \frac{1}{2}
\left( \bar\psi^+_2 \psi^+_2  |0\rangle_{RR} -
\bar\psi^+_1 \psi^+_1   |0\rangle_{RR}  \right)  \,.
\end{equation}
This motivates our ansatz for the boundary states 
$|B1\rangle\!\rangle$ and $|B2\rangle\!\rangle$: 
the ground state of $|B1\rangle\!\rangle$ couples to
$|\Psi_1^-\rangle$ (as well as to the states that correspond to
$(n=1)$ and ($n=3$)), while the ground state of $|B2\rangle\!\rangle$
couples to $|\Psi_2^-\rangle$ (as well as again to the states that
correspond to $(n=1)$ and ($n=3$)).  The corresponding D-branes should
therefore correspond to the (13)(24) and (14)(23) branes. As we have
shown in the main part, D-branes with these gluing conditions are
indeed obstructed under changing the Kaehler parameters of the two
tori separately.

\subsection{Spectral flow}

The above analysis implies that the orbifold RR ground states 
that correspond to the two polynomials $x_1^2 x_2^2$ and $x_3^2 x_4^2$
are precisely the states
\begin{equation} \label{a1}
|-,+\rangle \otimes \overline{|+,-\rangle} \qquad \hbox{and} \qquad
|+,-\rangle \otimes \overline{|-,+\rangle} \,.
\end{equation}
Indeed, the other two states 
$|+,+\rangle \otimes\overline{|-,-\rangle}$ and 
$|-,-\rangle \otimes \overline{|+,+\rangle}$ couple only to B-type
branes in the orbifold theory, and thus to A-type branes in the Gepner
model. 

On the other hand, the RR ground states corresponding to $x_1^2 x_2^2$
and $x_3^2 x_4^2$ must be the images under spectral flow of the NS-NS
sector states that describe the deformation of the 
Kaehler parameters of the two tori. These are the states  
\begin{equation} \label{a2}
\psi^{(1)}_{-1/2} \tilde{\bar\psi}^{(1)}_{-1/2} |0\rangle_{NSNS} \,,
\qquad 
\bar\psi^{(1)}_{-1/2} \tilde{\psi}^{(1)}_{-1/2} |0\rangle_{NSNS} 
\end{equation}
and
\begin{equation}
\psi^{(2)}_{-1/2} \tilde{\bar\psi}^{(2)}_{-1/2} |0\rangle_{NSNS} \,,
\qquad 
\bar\psi^{(2)}_{-1/2} \tilde{\psi}^{(2)}_{-1/2} |0\rangle_{NSNS} \,.
\end{equation}

As a final consistency check of our identification we can now show
that this is indeed the case. The spectral flow that defines a
symmetry of the Gepner model acts symmetrically on left- and
right-movers. In the torus orbifold, the corresponding flow should
therefore act asymmetrically (since in the identification mirror
symmetry has been performed). The first state in (\ref{a1}) has
$h_1=h_2=\bar{h}_1=\bar{h}_2=1/8$ and $q_1=-1/2$, $q_2=1/2$,
$\bar{q}_1=1/2$, $\bar{q}_2=-1/2$. Under spectral flow by one half
unit it therefore flows to a NS-NS state with the quantum numbers  
$h_1=1/2$, $\bar{h}_1=1/2$, $h_2=0$, $\bar{h}_2=0$, $q_1=-1$,
$\tilde{q_1}=1$, $q_2=0$, $\tilde{q_2}=0$. This is then precisely one
of the states in (\ref{a2}). The analysis for the other states is
similar.

\end{document}